\documentclass[sigconf,screen]{acmart}

\AtBeginDocument{%
  \providecommand\BibTeX{{%
    \normalfont B\kern-0.5em{\scshape i\kern-0.25em b}\kern-0.8em\TeX}}}

\usepackage[normalem]{ulem}
\usepackage{balance}
\usepackage{pifont}
\usepackage{hyphenat}
\usepackage{xurl}
\usepackage{textcomp}
\usepackage{microtype}

\usepackage[noend]{algpseudocode}
\usepackage{graphicx}
\usepackage{xcolor}
\usepackage[caption=false]{subfig}
\usepackage{caption}  
\usepackage{array}
\usepackage{enumitem}
\usepackage{multirow}

\usepackage{pifont}
\usepackage{soul}
\usepackage{comment}
\usepackage{listings}

\newcommand{\pname}[1]{{{{BaM}}}{#1}}

\newcommand{\fixme}[1]{{\textcolor{red}{TODO: \textit{#1}}}}
\newcommand{\vikram}[1]{{\textcolor{orange}{Vikram: \textit{#1}}}}

\newcommand{\zaid}[1]{{\textcolor{blue}{Zaid: \textit{#1}}}}

\newcommand{\reb}[1]{{\textcolor{black}{{#1}}}}
\newcommand{\modi}[1]{{\textcolor{black}{{#1}}}}

\newcolumntype{P}[1]{>{\centering\arraybackslash}p{#1}}

\definecolor{commentgreen}{RGB}{2,112,10}
\definecolor{eminence}{RGB}{108,48,130}
\definecolor{weborange}{RGB}{255,165,0}
\definecolor{frenchplum}{RGB}{129,20,83}
\definecolor{red}{RGB}{255,0,0}

\setcopyright{rightsretained}
\acmPrice{}
\acmDOI{10.1145/3575693.3575748}
\acmYear{2023}
\copyrightyear{2023}
\acmSubmissionID{asplosb23main-p88-p}
\acmISBN{978-1-4503-9916-6/23/03}
\acmConference[ASPLOS '23]{Proceedings of the 28th ACM International Conference on Architectural Support for Programming Languages and Operating Systems, Volume 2}{March 25--29, 2023}{Vancouver, BC, Canada}
\acmBooktitle{Proceedings of the 28th ACM International Conference on Architectural Support for Programming Languages and Operating Systems, Volume 2 (ASPLOS '23), March 25--29, 2023, Vancouver, BC, Canada}
\received{2022-07-07}
\received[accepted]{2022-09-22}

\begin{document}
\title{GPU-Initiated On-Demand High-Throughput Storage Access in the \pname{} System Architecture}


\author{Zaid Qureshi}
\authornote{Both authors contributed equally to this research.}
\authornote{Work was done while at UIUC.}
\email{zqureshi@nvidia.com}
\affiliation{\institution{NVIDIA/UIUC}
\country{USA}} 

\author{Vikram Sharma Mailthody}
\authornotemark[1]
\authornotemark[2]
\email{vmailthody@nvidia.com}
\affiliation{\institution{NVIDIA/UIUC}
\country{USA}} 

\author{Isaac Gelado}
\email{igelado@nvidia.com}
\affiliation{\institution{NVIDIA}\country{USA}} 

\author{Seungwon Min}
\authornotemark[2]
\email{davmin@nvidia.com}
\affiliation{\institution{NVIDIA/UIUC}\country{USA}} 

\author{Amna Masood}
\authornotemark[2]
\email{amnam2@illinois.edu}
\affiliation{\institution{AMDU/IUC}\country{USA}} 

\author{Jeongmin Park}
\email{jpark346@illinois.edu}
\affiliation{\institution{UIUC}\country{USA}} 

\author{Jinjun Xiong}
\email{jinjun@buffalo.edu}
\affiliation{\institution{University at Buffalo}\country{USA}} 

\author{C. J. Newburn}
\email{cnewburn@nvidia.com}
\affiliation{\institution{NVIDIA}\country{USA}} 

\author{Dmitri Vainbrand}
\email{dvainbrand@nvidia.com}
\affiliation{\institution{NVIDIA}\country{USA}} 

\author{I-Hsin Chung}
\email{ihchung@us.ibm.com}
\affiliation{\institution{IBM Research}\country{USA}} 

\author{Michael Garland}
\email{mgarland@nvidia.com}
\affiliation{\institution{NVIDIA}\country{USA}} 

\author{William Dally}
\email{bdally@nvidia.com}
\affiliation{\institution{NVIDIA/Stanford}\country{USA}} 

\author{Wen-mei Hwu}
\email{whwu@nvidia.com}
\affiliation{\institution{NVIDIA/UIUC}\country{USA}} 








\renewcommand{\shortauthors}{Qureshi, et al.}


\begin{abstract}
Graphics Processing Units (GPUs) have  
traditionally relied on the {host}  CPU to 
initiate access to the data storage.
This approach is well-suited for GPU applications with known data access patterns that enable partitioning of their dataset to be processed in a pipelined fashion in the GPU.
However, 
emerging applications such as graph and data analytics, recommender systems, or graph neural networks, require fine-grained, data-dependent access to storage.
CPU 
initiation of storage access is unsuitable for these applications due to high CPU-GPU synchronization overheads, I/O traffic amplification, and long CPU processing latencies. 
GPU-initiated 
storage 
removes these overheads 
from the storage control path and, thus,  
can potentially support these applications at much higher speed.
However,  
there is a lack of systems architecture and software stack that enable efficient 
GPU-initiated 
storage access. 
This work 
presents a novel system architecture, \pname{}, that
fills this gap. 
\pname{} 
features a fine-grained software cache to coalesce data storage requests while minimizing I/O traffic amplification. 
This software cache communicates with the storage system 
via high-throughput queues that enable the massive number of concurrent threads in modern GPUs to 
make I/O requests at a high
rate to fully utilize the storage devices and the system interconnect. Experimental results show that 
\pname{}  delivers 1.0$\times$ and 1.49$\times$ end-to-end speed up for BFS and CC graph analytics 
benchmarks while reducing hardware costs by up to 21.7$\times$ 
over 
accessing the graph data from the host memory. 
Furthermore, 
\pname{} speeds up data-analytics workloads by 5.3$\times$ over CPU-initiated storage access 
on the same 
hardware.

\end{abstract}





\maketitle

\section{Introduction}
\label{sec:intro}
After over a decade of phenomenal growth in compute throughput and memory bandwidth~\cite{fermi,amperewhite}, GPUs have become popular compute devices for HPC and machine learning applications.
Emerging high-value data-center workloads such as graph and data analytics~\cite{friendster,MOLIERE_2016,emogi, rapids}, graph neural networks (GNNs)~\cite{graphsage,obglsc}, and recommender
systems~\cite{dlrm,dlrmarch,massivedlrm, baidurecsys,dlrmscale} can potentially benefit from the compute throughput and memory bandwidth of GPUs. {These applications access massive datasets organized into array data structures 
whose sizes} range from tens of GBs to tens of TBs today, and are expected to grow rapidly in the foreseeable future.

Storing these datasets as in-memory objects enables applications to naturally and efficiently process the data. 
However, 
the memory capacity of GPUs, in spite of a 53$\times$ increase from that of G80 to A100~\cite{fermi, dgxa100dataset}, is only at 80GB, far smaller than the required capacity to accommodate entire datasets of these workloads.
Thus, some 
state-of-the-art approaches rely on CPU user/OS code to partition datasets into chunks and orchestrate the appropriate storage access and data transfers into the GPU memory for application processing.

Another approach is to rely on memory-mapped files and GPU page faults to active the CPU page fault handler to transfer data whenever the application 
accesses data not present in GPU memory.
In this paper, we refer to both alternatives as a {\bf CPU-centric approach}. 
Our experiments show that CPU-centric approaches 
are bottlenecked by CPU software and CPU/GPU synchronization overheads, resulting in poor performance (See $\S$~\ref{sec:eval} and~\cite{bamarxiv}).
To avoid the inefficiencies of the CPU-centric approaches, some state-of-the-art solutions use the host memory~\cite{emogi}, whose capacity typically ranges from 128GB to 2TB today, or pool together multiple GPUs' memories~\cite{dlrmscale} to hold the entire datasets. We 
refer to the use of host memory or pooling multiple-GPUs' memory to extend GPU memory capacity as a \emph{DRAM-only} 
solution. 

Extending the host memory to the level of tens of TBs is an extremely expensive proposition in the foreseeable future. Similarly, pooling 
multiple GPUs' memory to 
reach tens of TBs can 
be 
very expensive 
as well. 
The memory capacity of each A100 GPU is only 80GB, so a 10TB pool would require 125 A100 GPUs.
Furthermore, using the host memory and pooling GPU memories both require pre-loading the dataset into the extended memory. 
 Some of the preloaded data may not be ultimately used 
due to data-dependent accesses. 
We will 
show with results from graph analytics that the performance of a host-memory 
solution is comparable to or lower than our proposed approach despite its much higher cost.

\textbf{Proposal:} We propose a novel system architecture called \pname{} (Big accelerator Memory).
\pname{} capitalizes on the recent improvements in latency, throughput, cost, density, and endurance of storage devices 
to realize another level of the accelerator memory hierarchy.
The goal of \pname{}'s design 
is to provide efficient abstractions for the GPU threads to easily make on-demand, fine-grained accesses to massive datasets 
in the storage and achieve much higher application performance than state-of-the-art solutions.

\reb{Prior attempts\cite{activepointers,gpufs} to enable 
GPU threads to generate on-demand storage requests achieved 
low throughput ($\sim$823K IOPs on the NVIDIA A100 GPU), as shown in $\S$~\ref{sec:gds_gpufs_comp}.} In this paper, 
we present and evaluate the effectiveness of the key components and the overall design of \pname{} 
in addressing two 
key technical challenges in efficient 
on-demand storage accesses for accelerator applications.  

First, there is currently a lack of \textbf{fast} mechanisms for the GPU application code to generate on-demand storage access requests without incurring the CPU software bottlenecks such as the OS page fault handler. 
To fill this gap, ~\pname{} features a \modi{scalable,}  highly concurrent, high-throughput configurable software cache that takes advantage of the massive memory bandwidth and atomic operation throughput of modern GPUs.
\modi{The software cache coalesces redundant on-demand accesses and facilitates reuse of storage data while providing high application-perceived throughput.} 

Second, while the CPU-centric approaches 
suffer from the low-degree of CPU thread-level parallelism available to page fault handlers and device drivers, there is currently a lack of GPU mechanisms for orchestrating storage accesses \reb{without relying on the CPU}. 
To address this issue, ~\pname{} provides \modi{a user-level GPU} library of highly concurrent 
submission/completion queues in GPU memory that 
enables GPU threads whose on-demand accesses do not hit in the software cache to make storage accesses in a high-throughput manner. 
This user-level approach removes the page fault handling bottleneck, and
\modi{uses fine-grained synchronization and minimal critical sections to} 
reduce software overhead for each storage access and support a high-degree of thread-level parallelism. 

{There is a trend towards increasing autonomy and asynchrony of GPUs. The GPUDirect Async family~\cite{gpudirecttech} of technologies accelerate the control path when moving data directly into GPU memory from memory or storage. Each transaction involves initiation, where structures like work queues are created and entries within those structures are readied, and triggering, where transmissions are signaled to begin. 
{To our knowledge,} \pname{} is the first 
\textbf{\textit{accelerator-centric}} approach where GPUs can 
create
on-demand accesses to data where it is stored, be it memory or storage, {\bf without relying on the CPU to 
initiate 
or trigger 
the accesses.}  
Thus, \pname{} 
marks the beginning of a new variant of this family that is GPU kernel initiated (KI): \textit{GPUDirect Async KI Storage}.}

While the user-level 
storage device queues raise security concerns for traditional monolithic server architectures, the recent shift in data centers toward zero-trust security models that provide security guarantees through trusted hardware or software services
have provided the new system framework for securing accelerator-centric user-level storage access models like \pname{}~\cite{linefs, hyperlooppaper, DevFS}.
We have built a prototype~\pname{} system {through novel organization of} off-the-shelf hardware components and development of a novel custom software stack that takes advantage of the advanced architectural features in recent GPUs. 
Evaluation 
using a variety of workloads with
multiple datasets shows that~\pname{} is on-par with a 21.7$\times$ more expensive host-memory DRAM-only solution and up to \modi{5.3$\times$} faster than a state-of-the-art CPU-centric software solution. 
Overall, we make the following contributions. We 
\begin{enumerate}[topsep=0pt]
    \item propose~\pname{}, an accelerator-centric system architecture in which GPU threads perform 
    on-demand accesses to array data where it is stored, be it memory or storage\reb{, without relying on the CPU to initiate these accesses};
    \item enable on-demand, high-throughput fine-grained access to storage through \modi{a novel library of} highly concurrent {submission/completion protocol} queues;
    \item provide high-throughput, scalable software-defined caching and software API for programmers to exploit locality and control data placement for their applications; and
    \item {construct and evaluate a prototype design for GPUs to access massive storage data in a cost-effective manner.} 
\end{enumerate}




\reb{We have published a full version of the paper that includes appendix covering in-depth analysis on limitations of current system with additional evaluations~\cite{bamarxiv}. \pname{} is implemented completely in open-source, and both hardware and software requirements are publicly accessible~\cite{bamgithub, phdthesis1,phdthesis2}.}


\section{BACKGROUND}
\label{sec:background}
\reb{
This section covers the limitations with the common solution 
that keeps large datasets in abundant CPU memory or pooled multi-GPU memory ($\S$\ref{sec:cpumempool}) and then describe why GPUs can tolerate long storage access latency ($\S$\ref{sec:nvmeoverhead}).}

\subsection{Leveraging CPU or Pooled Multi-GPU Memory}
\label{sec:cpumempool}
Applications can leverage CPU memory or even pool together the memory of multiple GPUs to host large data structures. 
\modi{Previous works show that the GPU provides sufficient memory-level parallelism to 
tolerate the access latency of these memories and 
significantly out-perform the UVM~\cite{uvm} solution for 
graph traversal applications 
~\cite{emogi}.}

However, regardless of whether CPU memory or pooled GPU memory is used to host these data structures, this approach suffers from two major pitfalls.
First, \textit{data must still be loaded from the storage to the memory before any GPU computation can start}. 
Often this initial data loading can be the main performance bottleneck. 
(see the Target (T) system of Figure~\ref{fig:graphoverall}). 
Second, \textit{hosting the dataset in CPU memory or pooled GPU memory requires scaling the available memory}, by either increasing the CPU DRAM size or the number of GPUs in the system, with the dataset size, \textit{and thus can be {prohibitively expensive for massive datasets}}.

\subsection{
Tolerating Storage Access Latency}
\label{sec:nvmeoverhead}
As the storage device latency is reduced due to technological advancements like Optane~\cite{inteloptanewebsite} or Z-NAND~\cite{znand} media, software overhead is becoming a significant fraction of overall I/O access latency. 
\modi{Our experiments show that even with \texttt{io\_uring}, a highly optimized CPU software stack\cite{iouring}, \textit{as device latency decreases, OS kernel software overhead severely limits the storage access throughput and becomes a significant fraction, up to 36.4\%, of the total 
storage access latency.}}

To address this shift, emerging storage systems 
allow applications to make direct user-level I/O accesses to storage~\cite{wekafs, crossfs, ufs, zofs, spdk,DevFS,daos,Starta,splitfs}.
The storage system allocates user-level queue pairs, akin to NVMe I/O submission (SQ) and completion (CQ) queues, which the application threads can use to enqueue requests and poll for their completion.
Using queues to communicate with storage systems forgoes the userspace to kernel crossing of traditional file system access system calls.
Instead, isolation and other file system properties are provided through trusted services running as trusted user-level processes, kernel threads, or even storage system firmware running on the storage server/controller~\cite{wekafs,splitfs,crossfs,DevFS}. 

In such systems, the parallelism required to tolerate access latency and achieve full throughput of the device is 
governed by Little's Law: $T \times L = Q_d$, 
where $T$ is the target throughput, $L$ is the average latency, and $Q_d$ is the minimal queue depth required at any point in time to sustain the target throughput. 
To achieve the full potential of the critical resource, PCIe $\times$16 Gen4 connection providing $\sim$26GBps of bandwidth, then $T$ is $26GBps/512B = 51 M/sec$ and $26GBps/4KB = 6.35 M/sec$ for 512B and 4KB access granularities, respectively.
The average latency, $L$, depends on the SSD devices used {and is measured when the SSD provides its maximal throughput.} {For the experiments reported in this paper, $L$} 
is $11us$ and $324us$ for the Intel Optane and Samsung 980pro SSDs, respectively.
From Little's Law, to sustain a desired 51M accesses of 512B each, the system needs to accommodate a queue depth of  $51M/s \times 11us = 561$ requests (70 requests for 4KB) for Optane SSDs. 
For the Samsung 980pro SSDs, the required $Q_d$ for sustaining the same target throughput is $51M \times 324us = 16,524$ (2057 for 4KB). Note that $Q_d$ can be spread across multiple physical device queues. {To sustain $T$ over a computation phase, there need to be a substantially higher number of concurrently serviceable access requests than $Q_d$ over time}.

The emerging queue-based storage systems present major challenges to sustaining $T$ in massively parallel execution paradigms. 
Take as an example using NVMe queues to make I/O requests. 
After requests are enqueued into a NVMe SQ, the queue's doorbell must be rung with the updated queue tail to notify the storage controller of the new request(s).
As these doorbell registers are write-only, when a thread rings a doorbell, it must make sure that no other thread is writing to the same register and that the value it is writing is valid and is a newer value than any value written to that register before.
{Implementing queue insertion as a critical section, while simple, imposes significant} serialization latency, which might be fine for the CPU's limited parallelism but can cause substantial overhead for 
thousands of GPU threads.
Storage interfaces like \cite{DevFS,crossfs,splitfs} face similar serialization challenges.
\modi{These challenges are addressed in the design of the \pname{} queues.}

\section{\pname{} SYSTEM AND ARCHITECTURE}
\label{sec:design}

\begin{figure}[t]
\centering
{
  \includegraphics[width=\columnwidth]{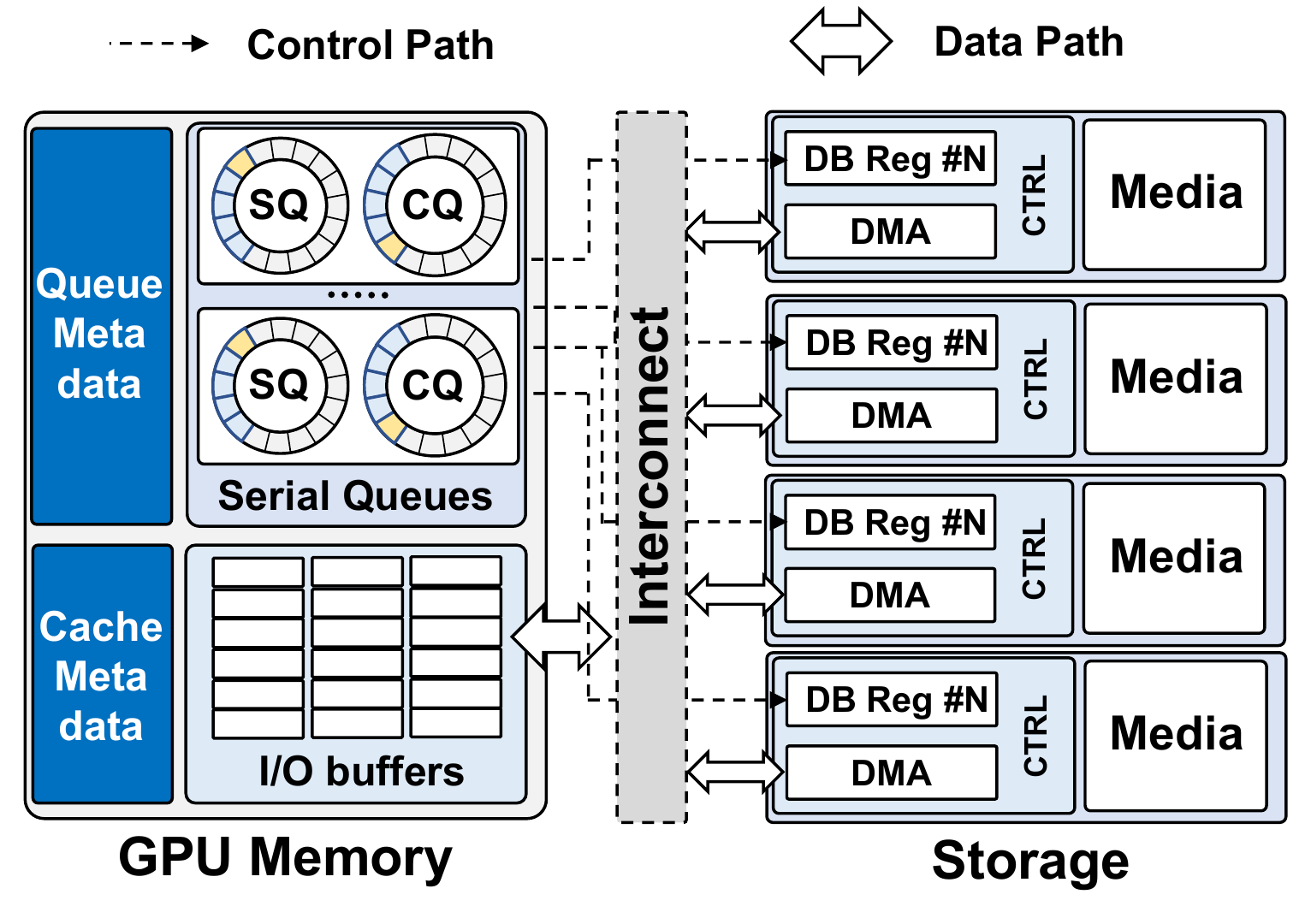}
}
\vspace{-5ex}
\caption{Logical view of~\pname{} design. 
}
\label{fig:logicalview}
\vspace{-4ex}
\end{figure}

The goal of \pname{}'s design is to 
provide high-level abstractions for accelerators to make on-demand, fine-grained, high-throughput access to storage while enhancing the storage access performance.
To this end, \pname{} provisions storage I/O queues and buffers in the GPU memory \modi{as shown in Figure~\ref{fig:logicalview}}, and builds on the recent memory-map capability of GPUs to map the storage doorbell registers to the GPU address space.
Although doing so enables the GPU threads to access terabytes of data on storage, \pname{} must address three key challenges in providing an efficient and effective solution:

\begin{enumerate}
   [leftmargin=*]
    \item  As storage protocols and devices exhibit significant latency, \pname{} must leverage the GPU's massive parallelism (up to three orders of magnitude more than the CPU)
    to keep many requests in flight, efficiently tolerate such latency, and {overlap it with computation}. (See $\S$\ref{sec:bam_io})
    \item As storage devices have {relatively low bandwidth} and GPUs have limited memory capacity, \pname{} must optimally utilize these resources. (See $\S$\ref{sec:bam_cache})
    \item As GPU kernels generally don't expect to make storage accesses, \pname{} must provide high-level abstractions that hide \pname{}'s complexity and make it easy for the programmers to integrate \pname{} into their GPU kernels. (See $\S$\ref{sec:bam_api})
\end{enumerate}

\noindent Next, we provide an overview of \pname{} and then explain how \pname{} addresses these challenges.

\subsection{\pname{} System Overview}
\label{sec:design:overview}

\pname{} presents
the \texttt{bam::array} high-level programming abstraction, enabling programmers to easily integrate \pname{} into their existing GPU applications.
An application can call \pname{} APIs to map the \texttt{bam::array} to data on storage, akin to \texttt{mmap}'ing a file.

\begin{figure}[t]
\centering
{
  \includegraphics[width=\columnwidth]{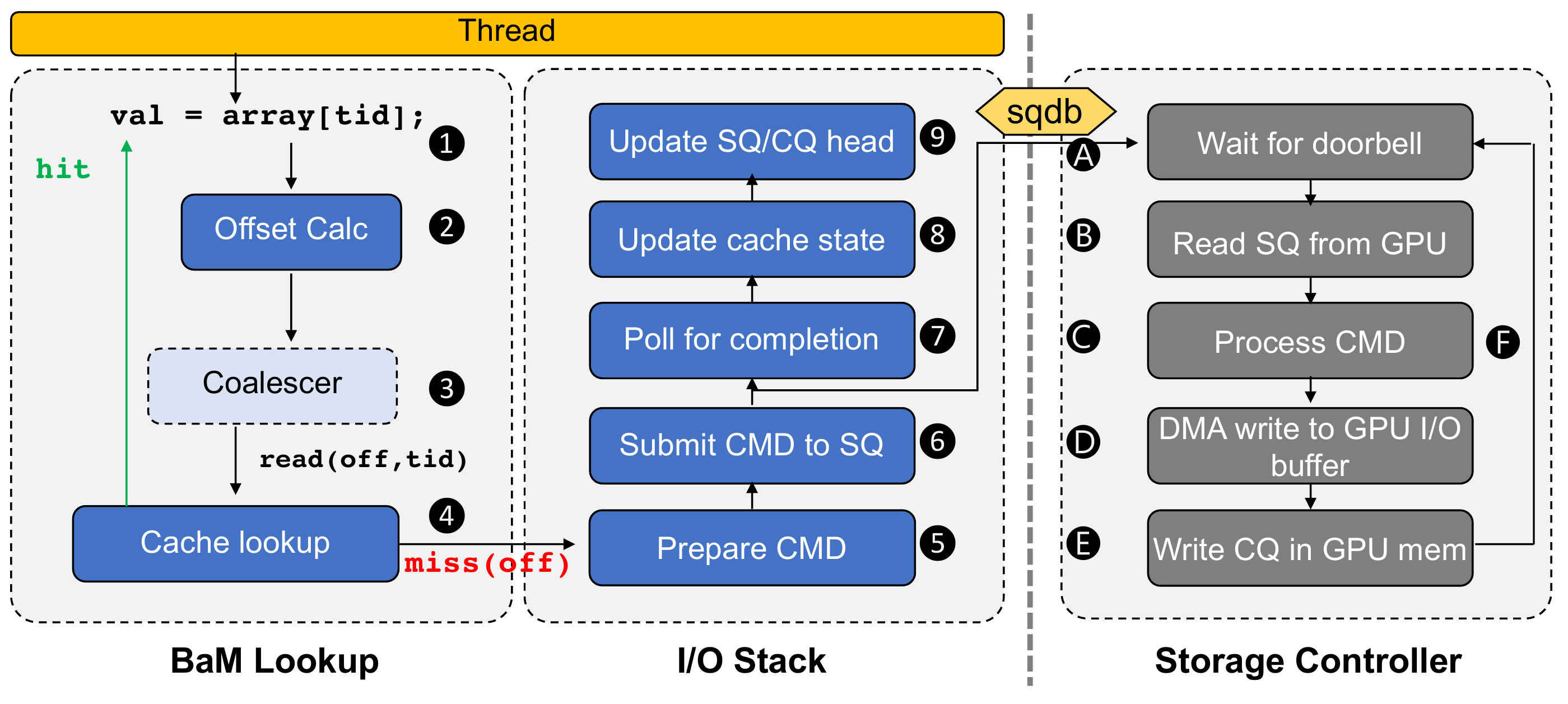}
}
\vspace{-3ex}
\caption{Life of a GPU thread in~\pname{}. 
}
\label{fig:bamdesign}
\vspace{-4ex}
\end{figure}

\modi{Figure~\ref{fig:bamdesign} shows how a GPU thread uses \pname{} to access data. 
When a GPU thread accesses data with the \texttt{bam::array} abstraction~\ding{182}, it uses the abstraction to determine the offset, i.e. cache line, for the data being accessed~\ding{183}. 
Threads in a warp can coalesce their accesses~\ding{184} if multiple threads access the same cache line.
For each unique cache line being accessed, a single thread probes the cache line's metadata~\ding{185} on behalf of the rest of the threads, improving cache access efficiency.}

If an access hits in the cache, the thread can directly access the data in GPU memory.
If the access misses, the thread needs to fetch data from the backing memory.
The \pname{} software cache is designed to optimize the bandwidth utilization to the backing memory in two ways: (1) by eliminating redundant requests to the backing memory and (2) by allowing users to \modi{configure the cache for their application's needs.}



If the storage system or device is backing the data, the GPU thread enters the \pname{} I/O stack to prepare a storage I/O request \ding{186}, enqueues it to a submission queue \ding{187}, and then waits for the storage controller to post the corresponding completion entry \ding{188}. 
The \pname{} I/O stack aims to amortize the software overhead associated with the storage {submission/completion} protocol by 
leveraging the GPU's immense thread-level parallelism, and enabling {low-latency} batching of multiple submission/completion (SQ/CQ) queue entries to minimize the cost of expensive doorbell register updates and reducing the size of critical sections in the storage protocol. 

On receiving the doorbell update (A), the storage controller fetches the corresponding SQ entries (B), processes the command (C) and transfer the data between SSD and the GPU memory (D). 
At the end of the transfer, the storage controller posts a completion entry in the CQ (E). After the completion entry is posted, the thread updates the cache state \ding{189}, 
update the SQ/CQ state \ding{190} and then can access the fetched data in GPU memory.

\subsection{Comparison With the  CPU-Centric Approaches}

\begin{figure*}[ht]
\centering
\subfloat[][CPU-Centric Model.]{
\includegraphics[width=0.46\linewidth]{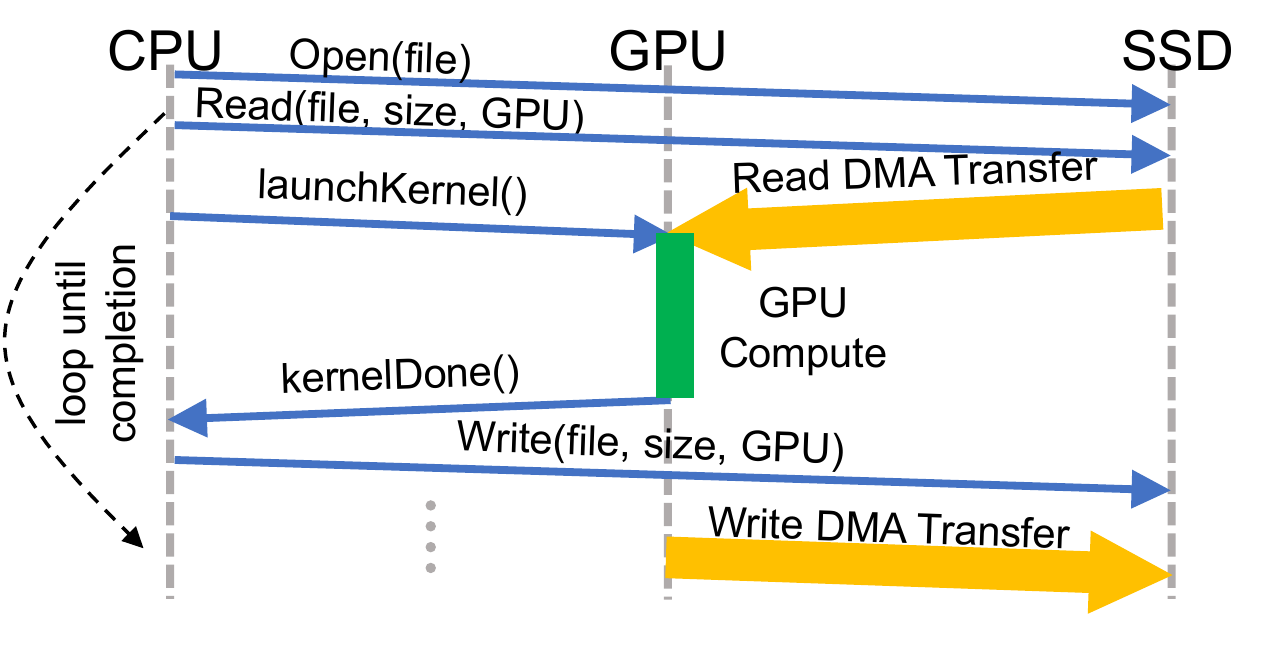}
\label{fig:cpucentrictimeline}}
\subfloat[][~\pname{} Model.]{
\includegraphics[width=0.45\linewidth]{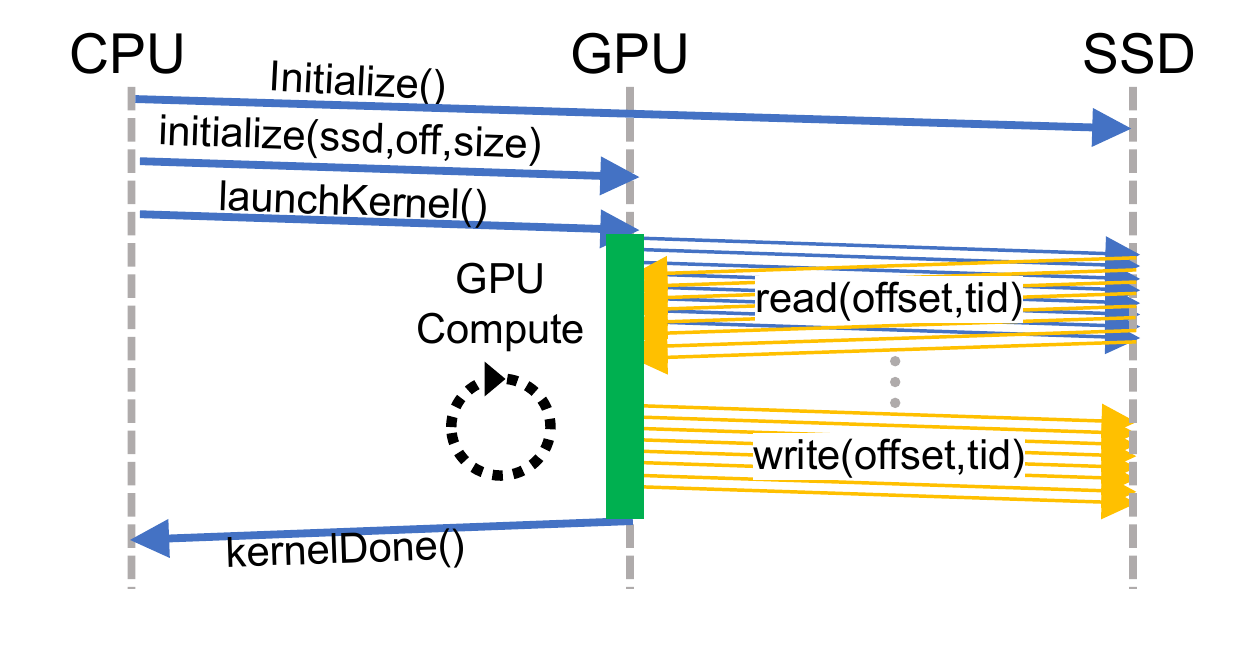}
\label{fig:gpucentrictimeline}}
\caption{Comparison between the traditional CPU-centric and \pname{} computation model is shown in (a) and (b).
{\pname{} enables GPU threads to directly access storage enabling compute and I/O overlap at fine-grain granularity.}
}
\label{fig:comparsion}
\end{figure*}
When compared to the proactive tiling CPU-centric approach shown in Figure~\ref{fig:cpucentrictimeline}, \pname{} has three main advantages.
First, \modi{with} proactive tiling, 
the CPU ends up copying data between the storage and {the} GPU memory and launching compute kernels multiple times to cover a large dataset. 
Each kernel launch and termination incurs costly synchronization between the CPU and the GPU.
\pname{} allows GPU threads to both compute and fetch data from storage, as shown in Figure~\ref{fig:gpucentrictimeline}, 
\modi{which reduces the frequency of CPU-GPU synchronization and GPU kernel launches.}
{Furthermore, the storage access latency of some threads can also be overlapped with the compute of other threads,which improves the overall performance.} 

Second, \modi{in proactive tiling,} because the compute is offloaded to the GPU and the data movement is orchestrated 
by the CPU, 
\modi{the CPU cannot accurately} determine which parts of the data are needed {and when they are needed, thus it ends} up fetching many unneeded bytes.
\modi{In contrast, with} \pname{}, a GPU thread 
\modi{fetches bytes only when they are used,} 
reducing the I/O amplification overhead.

Third, \modi{with} proactive tiling, programmers expend effort to partition the application's data and overlap compute with data transfers to hide storage access latency.
\pname{} allows the programmer to naturally {access the data through the array abstraction} and harness GPU thread parallelism across large datasets to hide the storage access latency.

\subsection{High-Throughput I/O Queues}
\label{sec:bam_io}
\pname{} leverages 
GPU's massive thread-level parallelism and fast hardware scheduling to maintain the queue depths needed to hide storage access latency and to 
achieve peak storage throughput.
However, ringing doorbells after enqueueing commands or cleaning up SQ entries, in the existing storage I/O protocols requires serialization.
{A critical section that encompasses the process of enqueuing a command and ringing the doorbell, albeit simple, would result in poor throughput and significant serialization latency when thousands of threads enqueue I/O requests concurrently.}
Instead, \pname{} uses fine-grain{ed} memory synchronization enabling many threads to enqueue into the SQ, poll the CQ, or mark queue entries for cleanup in parallel without \modi{large critical sections}.

We will use the SQ implementation in \pname{} as an example to explain how 
fine-grained memory synchronization helps to achieve the above design goals and principles. 
\pname{} keeps the following metadata per SQ in the GPU memory: 1) local copies of the queue's head and tail, 2) atomic \texttt{ticket} counter, 3) \texttt{turn\_counter} array, an integer array of the same length as the queue, 4) \texttt{mark} bit-vector 
of the same length as the queue, and 5) a \texttt{lock}.

{To enqueue requests, threads first atomically increment the ticket counter by two.}
The returned \texttt{ticket} value {is an index into a virtual queue with $2 ^ {32}$ entries and can be divided by the physical} queue size to assign each thread an \texttt{entry} in the {physical} queue, the remainder, and 
{a \texttt{turn} value along with a valid bit} 
{for that entry,} \modi{
the quotient
}. 
\modi{Using its \texttt{entry} and \texttt{turn}, each thread can wait for both the enqueue and dequeue operations of all previous commands in its assigned \texttt{entry} in the queue.}
Each thread uses its \texttt{entry} to index into the \texttt{turn\_counter} array, and polls on the location until the counter equals the thread's \texttt{turn}.
{That is, 
the \texttt{turn\_counter} array tracks the 
mapping of each physical queue entry to an virtual queue entry.} 
\modi{It effectively creates sub-queues whose waiting threads are ordered by their \texttt{turn} value} for each physical queue entry.

When it is the thread's turn, {i.e., the virtual queue entry assigned to the thread becomes active}, the thread can copy its I/O access command into its assigned (\texttt{entry}) position in the physical queue.
Afterwards, each thread sets the position's corresponding bit in the \texttt{mark} bit-vector.
The \texttt{turn\_counter} array enables as many threads as the size of the physical queue to 
copy their commands in parallel. 

The threads call the \texttt{move\_tail} routine to {complete the insertion of their requests. One thread will successfully acquire the lock and will move the tail past the consecutive new entries inserted by the calling threads and reset the \texttt{mark} bits associated with these entries with the \texttt{reset\_marks} routine. The thread then rings the queue's doorbell \modi{at the storage controller} with the new tail and releases the lock. \modi{This coalesces the doorbell writes, which are expensive operations over the PCIe interconnect, for multiple calling threads and improves the effective throughput.}
The rest of the calling threads will return if their \texttt{mark} bits are reset, signifying the SQ's tail will be moved past their enqueued entries.
If the {\texttt{mark}} bit for a thread has not been reset, the thread keeps trying to take the queue's \texttt{lock} and complete the insertion of its request.} 
\modi{Once the thread knows that its position's \texttt{mark} bit has been reset, it atomically increments the position's \texttt{turn\_counter} value {by one (to an odd value)}.}



After the thread’s command is submitted, the thread can poll, without any lock, on the CQ to find the completion entry for its submitted request. 
When it finds the completion entry, it 
{marks} the
\modi{entry for dequeuing by the next movement of the CQ head in the CQ's \texttt{mark} bit-vector.} 
The CQ’s head and doorbell are managed similarly as the SQ’s tail\modi{, except a thread 
{stops trying to reset its CQ entry's \texttt{mark} bit} if it notices the CQ's head has already been moved past the entry it marked for dequeueing}. 
{A winning thread among the calling threads rings the doorbell with a new \texttt{head} position to communicate forward progress to the storage system.}

The storage controller communicates forward progress by specifying 
{a} new SQ head in each CQ entry.
The thread with the CQ \texttt{lock} reads this field from the last CQ entry 
{whose} mark bit {was reset by it}
, and then iterates from the current SQ head until the specified new head, incrementing each position's \texttt{turn\_counter} value by one {(to an even value)}, allowing threads waiting for those positions to enqueue their commands.
The thread then updates the SQ head and releases the CQ \texttt{lock}. 


\subsection{\pname{} Software Cache}
\label{sec:bam_cache}
The \pname{} software cache is designed to enable optimal use of the limited GPU memory and 
storage-access bandwidth.
Traditional {OS-}kernel-mode memory management (allocation and translation) implementations must support diverse, legacy application/hardware needs. 
As a result, they contain large critical sections that limit the effectiveness of multi-threaded implementations. \pname{} addresses this bottleneck by allocating all the virtual and physical memory required for the software cache when starting each application. 
This approach 
\modi{reduces critical sections to only require} a lock when inserting or evicting a cache line, which allows the \pname{} cache to support many more concurrent accesses. 


The \pname{} cache aims to minimize the number of redundant I/O requests to the storage, avoiding unnecessary I/O traffic.
To this end, when a thread probes the cache with an offset, it checks the corresponding cache line's state.
If the accessed cache line is not in the cache, the thread locks the cache line, finds a victim to evict, and requests the cache line from the backing memory.
When the request completes, the requesting thread unlocks the cache line by making its state valid and incrementing its reference count.
This locking forces other threads that access the same cache line to wait until the cache line has been inserted and thus prevents multiple requests to the backing memory for the same cache line, exploiting locality and minimizing the number of requests to the backing memory. 
If, when the thread probes the cache line state, it is valid, the thread {atomically} increments the cache line's reference count.  
When the thread is done using
{the cache line, 
it decrements it's reference count.}

To avoid contention among concurrent evictions, the \pname{} cache uses a clock replacement algorithm~\cite{clockworkalgo}.
The cache has a global counter that gets incremented when a thread needs to find a cache slot.
The returned value of the counter assigns the thread a cache slot to use, allowing concurrent threads to evict unique cache slots in parallel.
If the assigned cache slot is currently mapped to a cache line that has a non-zero reference count, that is it is pinned, the thread will increment the counter again to attempt to replace another cache slot until it finds a cache slot that {is not} mapped to a pinned cache line. 
The thread will then mark the cache line invalid and change the mapping of the cache slot to point to the newly inserted cache slot.

\textbf{Warp Coalescing:} 
Threads in a warp may contend among themselves in accessing the \pname{} software cache, especially when consecutive threads try to access contiguous bytes in memory.
This contention incurs significant overhead when the needed cache line is already in the GPU memory.
To overcome this, \pname{}'s cache implements coalescing in software using 
the \texttt{\_\_match\_any\_sync} warp primitive to synchronize among threads in the warp and a mask is computed letting each thread know which other threads in the warp are accessing the same offset.
From that group, the threads decide on a leader and only the leader queries the cache and manipulates the requested cache line's state.
The threads in the group synchronize using the \texttt{\_\_shfl\_sync} primitive, and the leader broadcasts the address of the requested offset in the GPU memory to the group.

\reb{
Compared to warp coalescing implemented in prior works \cite{activepointers} where the unique cache probes in a warp are serialized and require many instructions, \pname{}'s coalescer enables all threads in a warp to divide themselves into groups according to the cache lines they are accessing with a single instance of the new \texttt{\_\_match\_any\_sync} warp synchronization primitive. 
A leader is determined for each group in parallel and each leader can independently probe the cache for the group's needed cache line, with no inter-group dependencies 
{or} synchronization.
}

\subsection{\pname{} Abstraction and Software APIs}
\label{sec:bam_api}

\modi{\pname{}'s software stack provides the programmer an array-based high-level API (\texttt{bam::array<T>}), consistent with array interfaces defined in modern programming languages (e.g. C++, Python, or Rust). 
As GPU kernels operate on such arrays, \pname{}'s abstraction 
{ minimizes} the programmer's effort to adapt their kernels, as shown in Listing~\ref{lst:bamrandom}.
\texttt{bam::array}'s overloaded subscript operator
enables the accessing threads to coalesce their accesses, query the cache, make I/O requests on misses, and returns the appropriate element of type T to the calling function.
The \texttt{bam::array<T>} can also be used to develop higher-level abstractions that can transparently provide optimizations like cache line reference reuse so threads do not need to probe the cache more than once when accessing the same cache line multiple times. 
In contrast, the proactive tiling CPU-centric model requires full, non-trivial application rewrites to  decompose the compute and data transfers into tiles that fit within the limited GPU memory.}


 
{\pname{} initialization requires allocating a few internal data structures that are reused during the application’s lifetime. 
Initialization can happen implicitly through a library construction if no customization is needed. 
Otherwise, the application specializes the memory through template parameters to \pname{}’s initialization call, a standard practice in C++ libraries. 
However, in most cases, specialization and fine-tuning are 
unnecessary, as we show later in $\S~\ref{sec:eval}$ where only \pname{}'s default parameters are used.
}



\begin{lstlisting}[basicstyle=\scriptsize,float=t!,linewidth=8.3cm,language=C++,caption={\small Example GPU kernel with \texttt{bam::array<T>}. 
%With  \texttt{bam::array<T>} abstraction, base GPU code requires changes only to the input arguments that are backed by \pname{} system.
\vspace{-2ex}}, label=lst:bamrandom ]
__global__ 
void kernel(bam::array<float> data, size_t n,
bam::array<float> out, bam::array<int> randidx) {
size_t tid = ...;
...
for(; tid < n; tid += (blockIdx.x * blockDim.x))
        out[tid] = data[randidx[tid]];
};
\end{lstlisting}



\section{\pname{} PROTOTYPE}
\label{sec:impl}
We use off-the-shelf hardware including NVIDIA GPUs and arrays of NVMe SSDs to {construct a \pname{} prototype} and show the benefits of allowing GPUs to directly access storage with enough random access bandwidth to {take full advantage of} a GPU's PCIe Gen4 x16 link. \textit{Once this level of data access bandwidth is achieved, a storage-based solution is as good as a host memory accessed through PCIe in terms of performance but much cheaper.} For simplicity, we describe the prototype assuming bare-metal, direct access to the NVMe SSDs.

\subsection{Enable Direct NVMe Access From GPU Threads}
\label{sec:gpudirect}
{For simplicity, we will use the NVMe SSD controllers as a simple example of storage controllers to explain the key features of the \pname{} prototype.} In order to enable GPU threads to directly access data on NVMe SSDs we need to: 
1) move the NVMe queues and I/O buffers from {the host} CPU memory to {the} GPU memory and 
2) enable GPU threads to write to the queue doorbell registers in the NVMe SSD's BAR space.
To this end, we create a custom Linux driver that creates a character device per NVMe SSD in the system, like the one by SmartIO~\cite{ssdgpunvme}. 
Applications use \pname{} APIs to \texttt{open} the character device for {each} SSD they wish to use.

In the custom Linux driver, \pname{} leverages GPUDirect RDMA APIs to pin and map NVMe queues and I/O buffers in the GPU memory.
This enables the SSD to perform peer-to-peer data reads and writes to the GPU memory. 

We leverage GPUDirect Async~\cite{gpudirecttech} to map the NVMe SSD doorbells to the CUDA address space so GPU threads can ring the doorbells on demand. This requires the SSD's BAR space to be first memory-mapped into the application's address space.
Then it is mapped to CUDA's address space with the \texttt{cudaHostRegister} API. 
Other storage systems can be enabled similarly.

\subsection{Scalable Hardware}
\label{sec:bam_hardware}
%


\modi{Scaling \pname{} using the PCIe slots available within a data-center grade 4U server 
is challenging as the number of PCIe slots available in these machines is limited. For instance, Supermicro AS-4124 system has five PCIe Gen4 $\times$16 slots per socket. If a GPU occupies a slot it can only access 4 $\times$16 PCIe devices without crossing the inter-socket fabric and suffering significant performance degradation.
However, as no single NVMe SSD can match the throughput of a PCIe x16 Gen4 link, the \pname{} hardware must scale the number of NVMe devices 
to provide the necessary throughput and match the bandwidth of the GPU's $\times$16 PCIe Gen4 interconnect.
}

{\renewcommand{\arraystretch}{1.2}
\begin{table}[t]
\centering
\scriptsize
    \caption{\small{\pname{} prototype system specification}}
    \begin{tabular}{|p{0.65in}|p{2.15in}|}
    \hline
    {\textbf{\pname{} Config}}   & {\textbf{Specification}} \\
    \hline
    \hline
    {System}      &  { Supermicro AS-4124GS-TNR}   \\
    \hline
    {CPUs}         &  { 2$\times$ AMD EPYC 7702 64-Core Processors }\\
    \hline
    {DRAM}         &  { 1TB Micron DDR4-3200 }    \\
    \hline
    {GPU}         &  { NVIDIA A100-80GB PCIe}    \\
    \hline
    {PCIe Expansion}   &  { H3 Platform Falcon-4016~\cite{h3}}   \\
    \hline
    {SSDs}         &  { Refer to Table~\ref{tab:ssdstat}}    \\
    \hline
    \multirow{1}{*}{Software}   &  { Ubuntu 20.04 LTS}, {NVIDIA Driver 470.82}, {CUDA 11.4} \\
    \hline

\end{tabular}
\label{tab:softhardconfig}
 \vspace{-2ex}
\end{table}
}

{\renewcommand{\arraystretch}{1.2}
\begin{table*}[t]
\centering
\scriptsize
    \caption{\small Comparison of different types of SSDs with DRAM DIMM. 
    {Prices are taken from a well known retailer~\cite{cdw} and near the time of writing this paper. The cost of required PCIe expansion hardware~\cite{h3} is included for systems using SSDs.}}
 \begin{tabular}{|P{0.67in}|P{0.3in}|P{0.8in}|P{.97in}|P{.93in}|P{0.65in}|P{0.3in}|P{0.23in}|P{0.18in}|}
     \hline
    {\textbf{Technology}}& {\textbf{Sources}} & {\textbf{Product}}  & {\textbf{RD IOPs (512B, 4KB)}}& {\textbf{WR IOPs (512B, 4KB)}} & {\textbf{Latency ($\mu$s)}} & {\textbf{DWPD}}  & {\textbf{\$/GB}} & {\textbf{Gain}}  \\
    \hline
    \hline
	{DRAM}                                   & multiple & DIMM (DDR4)     & $>$10M                & $>$10M      & O(0.1) & $>$1000 & 11.13 & 1.0$\times$ \\
    \hline
	{Optane~\cite{inteloptanewebsite}}     & single  & Intel P5800X     & 5.1M, 1.5M            &  1M, 1.5M   & O(10) &   100    &  2.54 & 4.4$\times$ \\
    \hline
	{Z-NAND~\cite{znand}}                   & single   & Samsung PM1735    & 1.1M, 1.6M            & 351K, 351K  & O(25) &  3       &  2.56 &4.3$\times$ \\
    \hline
    {NAND Flash \cite{980pro}}            & multiple  & Samsung 980pro   & 700K-800K, 700K-800K   & 172K, 172K  & O(100)  &  0.3   &  0.51 & 21.8$\times$ \\
    \hline
\end{tabular}
\label{tab:ssdstat}
\end{table*}
}

To address this, we built a custom prototype machine for the \pname{} architecture using the off-the-shelf components as shown in 
Table~\ref{tab:softhardconfig}. 
The \pname{} prototype uses a PCIe expansion chassis with custom PCIe topology for scaling SSDs. 
The PCIe switches provide low-latency, high-throughput peer-to-peer access.
The expansion chassis has two identical drawers that can be connected independently to the host.  
Each drawer supports 8 $\times$16 PCIe slots.
We use one $\times$16 slot in each drawer for an NVIDIA A100 GPU and the rest of the slots are populated with different types of SSDs.
Currently, each drawer can only support 10 U.2 SSDs as they take significant space.
With PCIe bifurcation, a multi-SSD riser card can enable more than 16 M.2 SSDs per drawer. 

\textbf{SSD Technology trade-offs:} 
\modi{Table~\ref{tab:ssdstat} lists the metrics that significantly impact the design, cost and efficiency of \pname{} systems for three types of off-the-shelf SSDs. 
The \texttt{{\$/GB}} 
is based on the current list price per device, the expansion chassis, and the risers needed to build the system. 
A comparison of these metrics across SSD types shows that the consumer grade NAND Flash SSDs are inexpensive with more challenging characteristics, while the low-latency drives such as Intel Optane SSD and Samsung Z-NAND are more expensive with desirable characteristics. 
For example, for write intensive applications using \pname{}, Intel Optane drives provide the best througput and endurance.}

\textit{Irrespective of the underlying SSD technology, as shown in Table~\ref{tab:ssdstat}, the \pname{} prototype provides 
4.3-21.8$\times$ advantage in cost-per-GB, even with the expansion chassis and risers, over a DRAM-only solution. 
Furthermore, this advantage grows with additional capacity added per device, which makes \pname{} highly scalable as SSD capacity and application data size increase.  
}

\subsection{\pname{} Raw Throughput}
\label{sec:eva:micro}
We establish that \pname{} can generate sufficient I/O requests to saturate the underlying storage system by measuring raw throughput of \pname{} using microbenchmarks with Intel Optane SSDs 
and an NVIDIA A100 GPU.
We allocate all the available SSD SQ/CQ queues into {the} GPU memory with a queue depth of 1024. 
We then launch a CUDA kernel with each thread requesting a random 512-byte block from the SSD via a designated queue. 
The requests are uniformly distributed  across all queues with round-robin scheduling.
We vary the number of threads and SSDs mapped to the GPU.
For multiple SSDs, the requests are uniformly distributed across SSDs using round-robin scheduling. 
We measure I/O operations per second (IOPs) as the ratio of the number of requests submitted and the kernel execution time.

\begin{figure}[t]
\centering
\subfloat{
\includegraphics[width=0.40\textwidth]{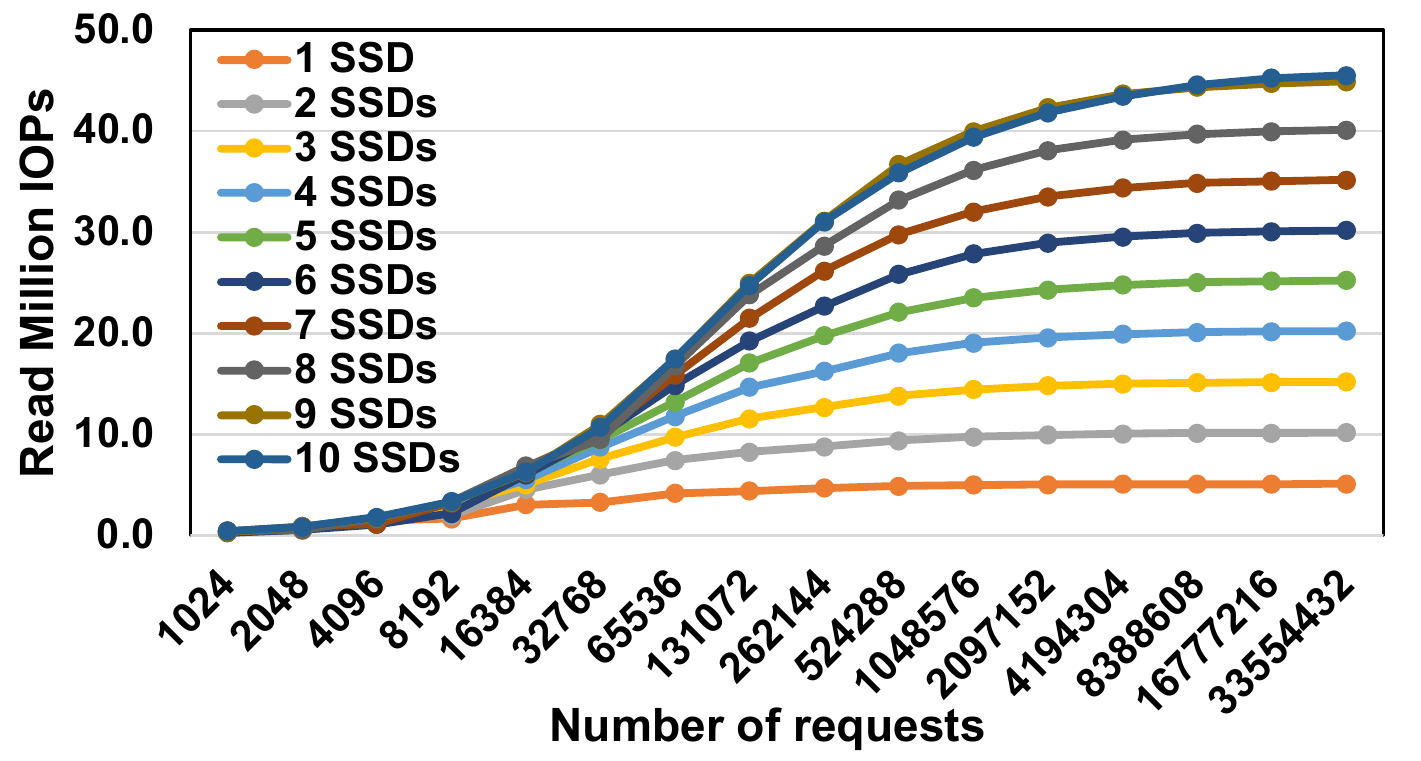}
\label{fig:randread}}
\qquad
\subfloat{
\vspace{-6ex}
\includegraphics[width=0.40\textwidth]{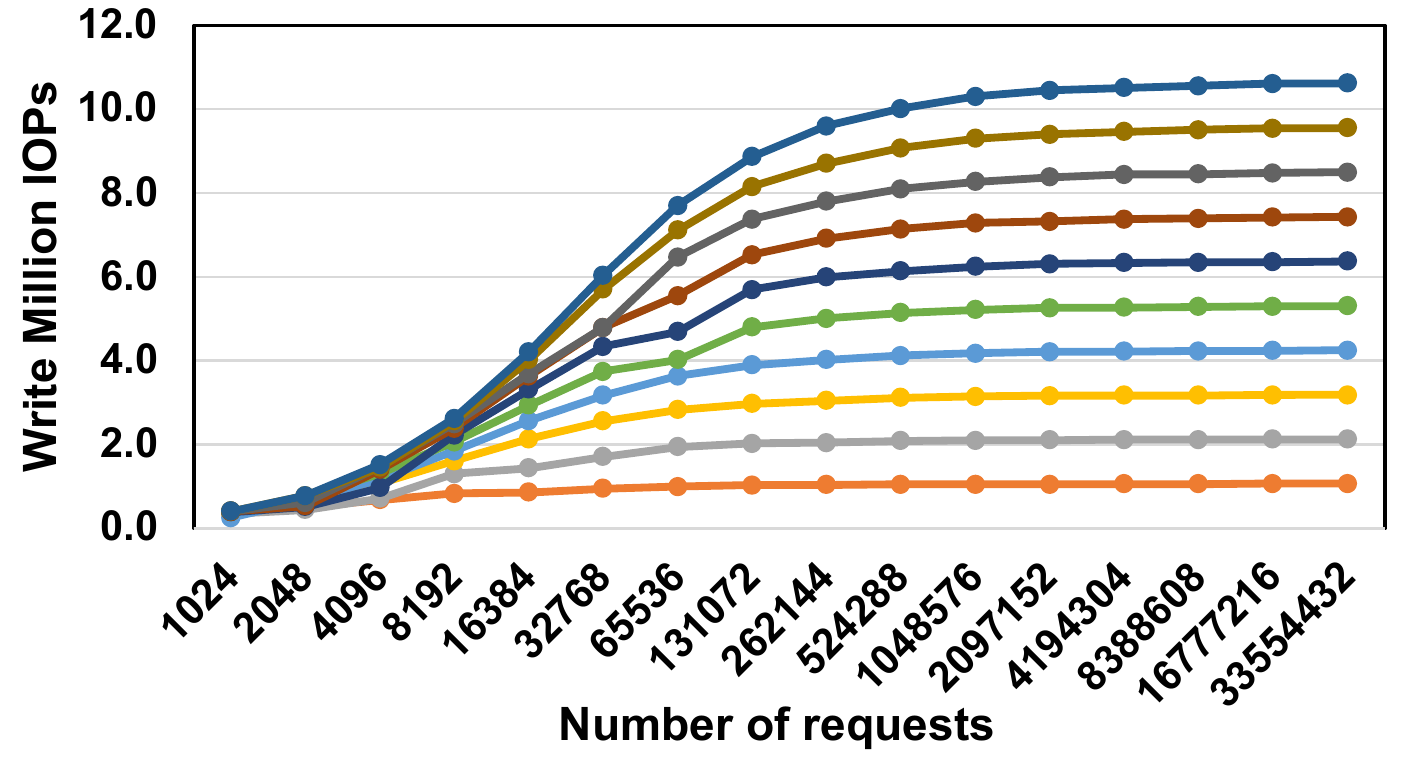}
\label{fig:randwrite}}
\caption{512B random read (top) and write (bottom) benchmark scaling with \pname{} on Intel Optane P5800X SSDs. 
{\pname{}'s I/O stack can reach peak IOPs per SSD and linearly scale for random read and write accesses.}
}
\label{fig:microbench}
\vspace{-4ex}
\end{figure}

\textbf{Results:}
Figure~\ref{fig:microbench} presents the measured IOPs for 512B random read and write benchmarks.
\pname{} can reach peak IOPs per SSD and linearly scale with  additional SSDs for both reads and writes.
With a single Optane SSD, \pname{} only requires about 16K-64K GPU threads to reach near peak IOPs (see Table~\ref{tab:ssdstat}). 
With ten Optane SSDs, \pname{} achieves 45.8M random read IOPs and {10.6M} random write IOPs, the peak possible for 512B accesses to the Intel Optane SSDs.
This is 22.9GBps \modi{(90\% of the measured peak bandwidth for Gen4 $\times 16$ PCIe links)} and 5.3GBps of random read and write bandwidth, respectively. Further improvements in 
write bandwidth, which isn't yet hitting PCIe limits, can be 
\modi{achieved} by scaling to more SSDs. 
Similar performance and scaling is observed with Samsung SSDs and also at 4KB access sizes 
but are not reported here due to space constraints.
\textit{These results validate that \pname{}'s {infrastructure software} can match the peak performance of the underlying storage system.}

\subsection{\reb{Discussion}}
\reb{
\textbf{GPUDirect RDMA I/O Consistency:}
As prior works\cite{gtc_rdma, lynx} have noted, when a third-party device writes into the GPU's memory with GPUDirect RDMA over PCIe, without a PCIe read following the writes, the order of these writes might not be preserved from the viewpoint of the GPU threads 
{running concurrently}.
In the \pname{} prototype, we allow a GPU thread 
to submit a second I/O request after successfully polling for the first one's completion.
As the storage device must read a submitted request over PCIe before writing the completion entry for it, when the thread finds the completion entry for the second command, it knows a PCIe read 
{has been completed after all} 
the writes for the first command. 
However, doing so incurs a 100\% performance overhead.
}

\reb{
To reduce the total number of additional I/O requests submitted, we implement a shared global virtual queue with 
a \texttt{lock}.
{All threads who find the CQ entries for their commands race for the \texttt{lock} of the virtual queue,} and the winner enters the \pname{} I/O stack to submit an additional I/O request {on behalf of all competing threads whose request can be coalesced with its own}. 
After the winner finds the completion entry for the second 
{request,} it 
{notifies the threads whose second requests were coalesced and} releases the \texttt{lock}.
Other threads continuously poll to see if {their {respective} second request{s are} 
covered through coalescing.}
{If} not they try to obtain the \texttt{lock} again.
We find that this solution 
{incurs only} less than 8\% performance overhead. 
}

\reb{
\textbf{Programming Model:}
\pname{} provides the GPU threads a memory-like abstraction, with the same memory consistency {properties} 
as {the} GPU memory. If the application threads read and write to the same words in \pname{}-backed memory, then it is up to the application to implement the necessary synchronization to avoid races, just as if the threads were reading and writing to {the} GPU memory. 
}

\reb{
\pname{} supports writes at all levels of its software stack: block write I/O requests, tracking dirty cache lines, and the write method in the high-level abstractions. When the high-level abstraction’s write method is called, the access to the cache line’s state will set the dirty bit. The \pname{} cache is a write-back cache and has APIs for the user to flush a specific or all dirty cache lines.
}

\reb{
If the system crashes in the middle of a GPU kernel that writes, there are no guarantees unless the application itself takes the responsibility of check-pointing application state and data{, which} 
is common practice in GPU accelerated applications.
}

\reb{
\textbf{CPU-GPU Data Sharing:}
{If} the application instantiates a \pname{} cache in {the} GPU memory and another \pname{} cache in {the host} CPU memory,
it is up to the application to implement the appropriate synchronization and communication to maintain a consistent view of shared data, just like when the application uses GPU memory without \pname{}.
}

\section{EVALUATION}
\label{sec:eval}

\pname{} excels on applications that have data-dependent data access patterns.  
Data dependent access patterns are widely used 
in popular applications, e.g. graph analytics, and frameworks, e.g. RAPIDS for accelerating analytical queries. 
In this section, we present an evaluation of the prototype \pname{}
system using these two workloads with a variety of datasets and show that 
a) \pname{}'s performance is either on-par with or outperforms the state-of-the-art solutions (see $\S$~\ref{sec:evalgraph} and $\S$~\ref{sec:eval:ioamp}). 
b) \pname{}'s design is agnostic to the SSD storage medium used, enabling application-specific cost-effective solutions. 
c) \pname{} reduces I/O amplification and CPU orchestration overhead significantly for data-analytics workloads (see $\S$~\ref{sec:eval:ioamp}).

\begin{figure}[t]
\centering
{
  \includegraphics[width=\columnwidth]{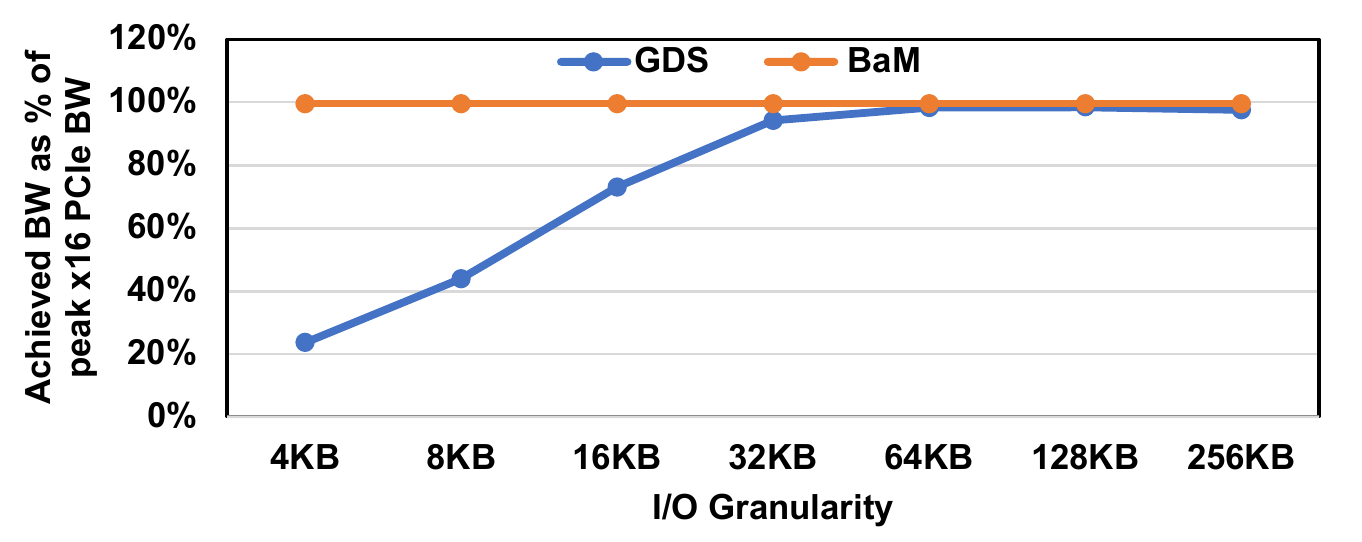}
}
 \vspace{-3ex}
\caption{\reb{Performance of \pname{} compared with NVIDIA GDS~\cite{gds}. With granularities less than 32KB, GDS is not able to saturate the PCIe interface due to the overheads of the traditional CPU software stack. In contrast, \pname{} is able to saturate the interface ($\sim$25GBps) at even 4KB I/O granularity.}}
\label{fig:gdscompare}
\end{figure}

\reb{\subsection{Comparison With GDS and ActivePointers}\label{sec:gds_gpufs_comp}}
\reb{We evaluate \pname{}'s performance benefits over NVIDIA GDS \cite{gds} for different I/O block sizes, ranging from 4KB to 1MB.
For GDS, we use \texttt{fio} to benchmark sequential access performance to transfer 128GB of data from four SSDs to GPU memory with 16 CPU threads. 
In the case of \pname{}, each warp is assigned a cache-line, the same size as the I/O blocks used for GDS, and consecutive warps access consecutive cache-lines in the 128GB dataset.
Recall that in \pname{} the cache-line size defines the I/O access granularity.
As shown in Figure~\ref{fig:gdscompare}, GDS can only saturate the GPU's PCIe link at the large I/O granularity of 32KB and only reaches 23.6\% of the PCIe bandwidth at 4KB.
Regardless of the number of CPU threads used, GDS is limited by the high overhead incurred by the Linux software stack.
In contrast, \pname{} {easily achieves ~25GBps using four SSDs, which is the measured peak bandwidth of the GPU's PCIe link.}
}

\begin{figure}[t]
\centering
{
  \includegraphics[width=\columnwidth]{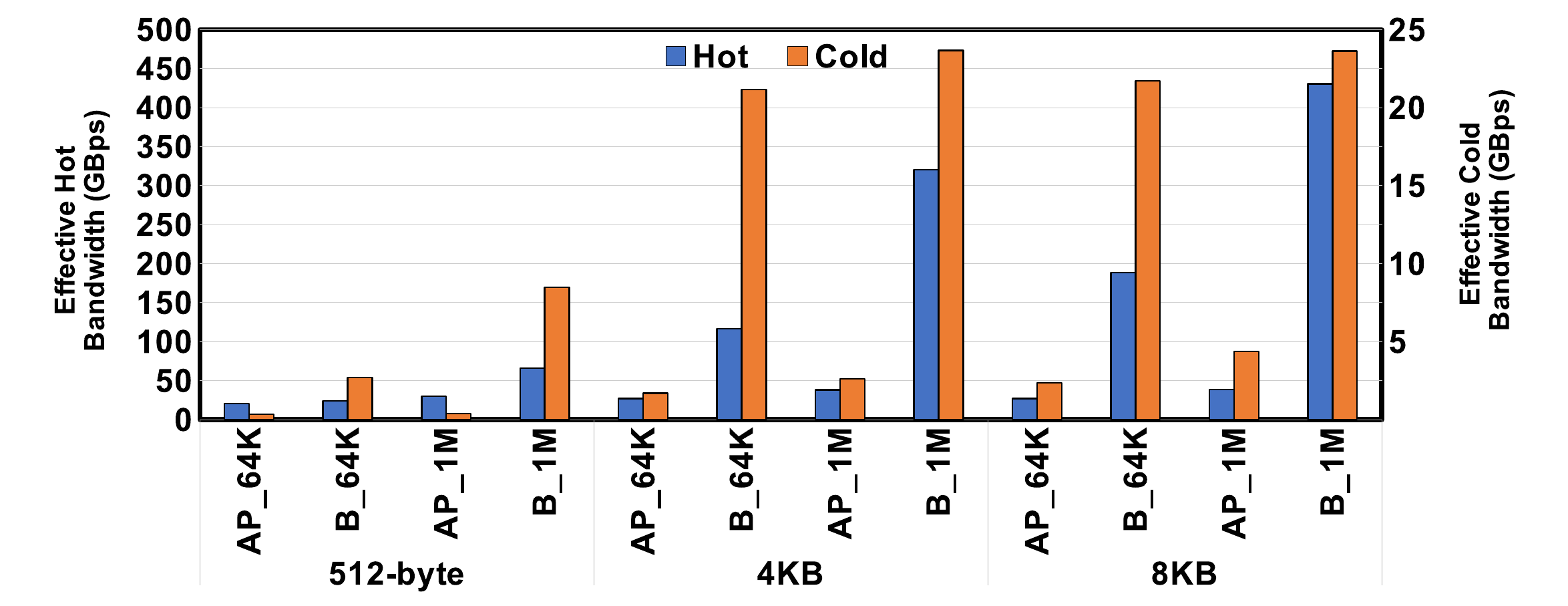}
}
 \vspace{-2ex}
\caption{\reb{Performance of \pname{} (\texttt{\textbf{B}}) and ActivePointers+GPUfs~\cite{activepointers,gpufs} (\texttt{\textbf{AP}}) with 64K and 1Million GPU threads with a 8GB cache in the GPU memory, in both hot and cold state, using 512-byte, 4KB, and 8KB cache-line sizes. 
\pname{} provides a peak miss-handling throughput of 17MIOPs with 4 Optane SSDs, which is 20.7$\times$ higher than ActivePointers' peak miss-handling throughput with the fast CPU memory. 
\pname{}'s provides a peak hot cache bandwidth of 430GBps, 11.2$\times$ higher than ActivePointers' cache.
} 
}
\label{fig:gpufscompare}
\vspace{-2ex}
\end{figure}

\reb{Next, we compare \pname{} and ActivePointers~\cite{activepointers}.  
For this evaluation, a warp is assigned to read 1024 contiguous 8-byte elements in a file where 
threads access the elements in a coalesced manner.  
With ActivePointers, the file is pinned in the Linux page cache in CPU memory, favoring ActivePointers as \textit{any misses in the ActivePointers cache only require data transfers from CPU memory to GPU memory, avoiding storage I/O requests and latencies}.
For \pname{}, the data is kept on four SSDs from where it is requested in case of a miss in the \pname{} cache, incurring storage access latency.  
Due to space constraints{,} we only show the results for 64K and 1 Million threads in Figure~\ref{fig:gpufscompare}; {measurements with up to} 32 Million threads show similar trends.
}

\reb{With a cold cache, ActivePointers' I/O performance is greatly hindered by the GPUfs~\cite{gpufs} mechanism that GPU threads use to request the data transfers from the CPU, achieving only 
an effective bandwidth of 4.4 GBps for 8 KB data transfers out of CPU memory.
ActivePointers provides a peak miss-handling throughput of 823 KIOPs, with 512-byte cache-lines.
In contrast,  \pname{} nearly saturates the GPU's PCIe link at around ~24 GBps with 4 KB and 8 KB cache-lines. Even with 512-byte cache-lines, \pname{} achieves 85\% of the peak throughput supported by the 4 SSDs, at 17 MIOPs. 
Furthermore, when both caches are hot, \pname{} can provide an effective bandwidth of up to 430 GBps, 11.2$\times$ more than ActivePointers' peak bandwidth.
{\pname{} outperforms ActivePointers by more than an order of magnitude in both cache miss handling and hit data delivery.} 
}

{\renewcommand{\arraystretch}{1.2}
\begin{table}[t]
\centering
\scriptsize
    \caption{\small Graph Analytics Datasets.  
   }
  \label{tab:graphdataset}
  \begin{tabular}{|P{1in}|P{0.51in}|P{0.50in}|P{0.40in}|}
    \hline
    \textbf{{Graph}} &  \textbf{{Num. Nodes} }& \textbf{{Num. Edges}}  & \textbf{Size (GB)} \\
    \hline
    \hline
    GAP-kron (\textbf{\texttt{K}})       \cite{GAP}                  & 134.2M   & 4.22B  & 31.5 \\
    \hline
    GAP-urand (\textbf{\texttt{U}})      \cite{GAP}                  & 134.2M   & 4.29B  & 32.0 \\
    \hline
    Friendster (\textbf{\texttt{F}})    \cite{friendster}           & 65.6M    & 3.61B  & 26.9 \\
    \hline
    MOLIERE\_2016 (\textbf{\texttt{M}})  \cite{MOLIERE_2016}         & 30.2M    & 6.67B  & 49.7 \\    
    \hline
    uk-2007-05 (\textbf{\texttt{Uk}})     \cite{BoVWFI,BRSLLP}        & 105.9M   & 3.74B  & 27.8\\
    \hline
  \end{tabular}
\end{table}
}

\subsection{Performance Benefit for Graph Analytics}
We evaluate the performance benefit of \pname{} for graph analytics applications. 
We use the graphs listed in Table~\ref{tab:graphdataset} for the evaluation. K, U, F, M are the four largest graphs from the SuiteSparse Matrix collection~\cite{suitesparse} while the Uk 
is taken from LAW~\cite{ubicrawler}. 

A goal of \pname{} is to provide competitive performance against the host-memory-based DRAM-only graph analytics solution. 
\modi{To this end, \textit{optimistic} target baseline 
\textbf{\texttt{T}}
allows the GPU threads to directly perform coalesced fine-grain access to the graph data stored in the host-memory 
~\cite{emogi}.
As there is sufficient CPU memory for the input graphs used in this experiment, we can make a direct performance comparison 
between \pname{} and \textbf{\texttt{T}}.}

We run two graph analytics algorithms, Breadth-first-search (BFS) and Connected Components (CC), on the target system and \pname{} with different SSDs listed in the Table~\ref{tab:ssdstat}.
Porting the state-of-the-art GPU implementations for both applications~\cite{emogi} to \pname{} requires minimal code changes as described in $\S$\ref{sec:bam_api}.
For BFS, we report the average run time after running at least 32 source nodes with more than two neighbors.
We do not execute CC on the Uk dataset since CC operates only on undirected graphs.
Finally, we fix the \pname{} software cache size to 8GB, the cache-line size to 4KB and 
\modi{unless explicitly stated, we use four Intel Optane SSD with 128 queue-pairs at 1024 depth.}

\label{sec:evalgraph}

\begin{figure*}[t]
\centering
{
  \includegraphics[width=\textwidth]{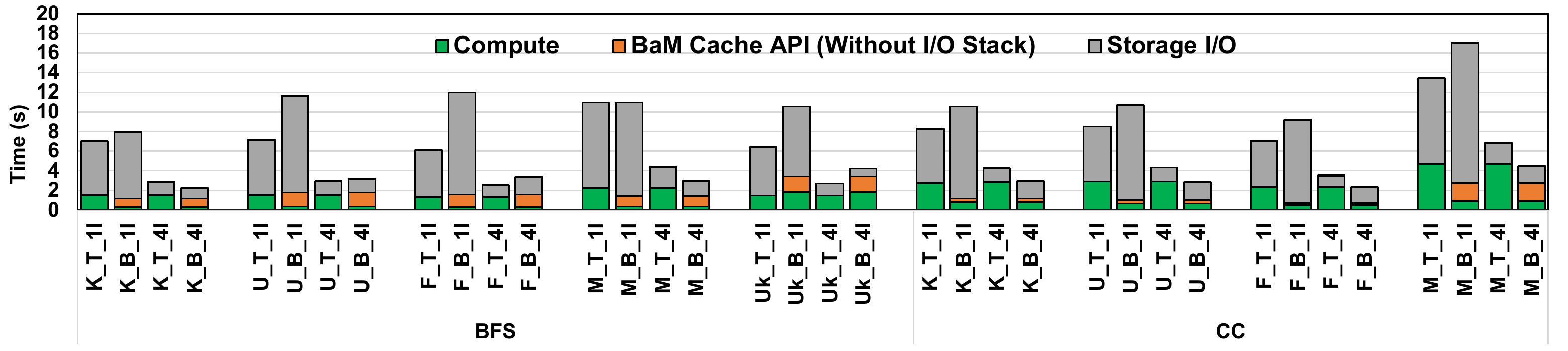}
}
\vspace{-4ex}
\caption{\reb{Graph analytics performance of \pname{} (CL size 4KB) and the Target (T) system with a single Intel Optane SSD. 
{On average, \pname{}'s end-to-end time is 1.0$\times$ (BFS) and 1.49$\times$ (CC) faster than the Target.}}}
\label{fig:graphoverall}
\end{figure*}

\textbf{Overall performance with Intel SSD:}
Figure~\ref{fig:graphoverall} shows the performance of both the target system (\textbf{\texttt{T}}) and \pname{} with single 
{(\texttt{B\_1I})} and four 
{(\texttt{B\_4I})} Intel Optane SSDs. 
For the single SSD 
{(\texttt{B\_1I})} configuration, 
\pname{} is on average slower by 1.43$\times$ and 1.27$\times$ for BFS and CC workload, respectively.
This is because of the limited storage throughput available for \pname{} with single SSD ($\times$4 PCIe Gen4 interface).

\modi{Scaling the number of SSDs to four 
{(\texttt{B\_4I})} and replicating data increases the \pname{}'s aggregate bandwidth and}
provides similar bandwidth as the $\times$16 Gen4 PCIe interface as in the target system.
Comparing with the target system \textbf{\texttt{T}} with file-loading time, \textit{\pname{} with four Optane drives provides on average \modi{1.00$\times$} and \modi{1.49$\times$} speedup on BFS and CC applications, respectively.}

In both workloads, \pname{}
overlaps the SSD data transfers for some threads with the compute of other threads, {fully utilizing 
PCIe Gen4 $\times$16 bandwidth while incurring}
much less I/O amplification.  
In contrast, the target system \textbf{\texttt{T}} needs to wait until the file is loaded into memory before it can offload the compute to GPU. The superior host-memory bandwidth of the \textbf{\texttt{T}} system
cannot overcome the initial file loading latency. As a result, \pname{} achieves either on-par or higher end-to-end performance.


Compared to the single-SSD \pname{} configuration, the four-SSD configuration scales 
to \modi{3.48$\times$ for BFS and 4$\times$} for CC. 
\modi{The main detractor of scaling for BFS is the \texttt{Uk} dataset. 
{Many nodes in the \texttt{Uk} graph have tiny neighborlists, resulting in deep BFS traversal ($>$100+ iterations) with small frontiers. Consequently, the total number of overlapping I/O requests
in each iteration of BFS 
is insufficient to tolerate the latency.}
}


\begin{figure}[t]
\centering
{
  \includegraphics[width=\columnwidth]{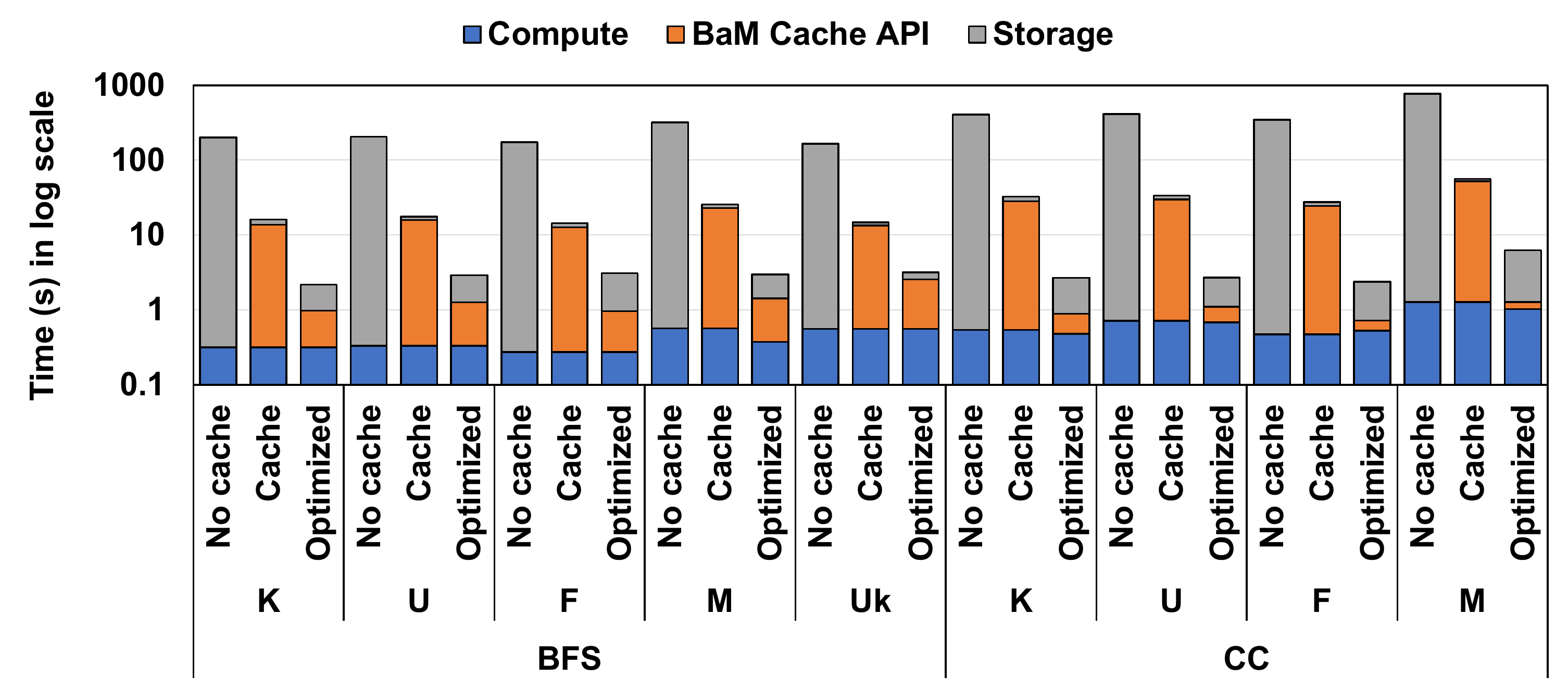}
}
\vspace{-4ex}
\caption{\reb{Sources of performance improvement in \pname{}. Adding a naive cache in \pname{} provides on average 11.9$\times$ (BFS) and 12.65$\times$(CC) speedup over not having a cache. When the application leverages and exploits \pname{}'s warp coalescing and reference reuse efficiently (\texttt{Optimized}), the application is sped up further by 6.07$\times$ (BFS) and 11.24$\times$ (CC) on average.}}
\label{fig:sourceperf}
\vspace{-2ex}
\end{figure}

\textbf{\pname{} performance breakdown:}
We now slice up the overall execution time of \pname{} to understand how each \pname{} component contributes to the overall execution time, \modi{as shown in Figure~\ref{fig:graphoverall}.}
We first load the entire dataset in the GPU-HBM memory and measure the execution time (\texttt{Compute} in green color). 
Next, we measured the total execution time when all the data is in the GPU-HBM memory, but each application access must go through the \pname{} cache API. 
This captures the best case performance one can achieve with \pname{} cache with no I/O requests. 
Subtracting \texttt{Compute} time from this measured execution time provides the cache API overhead (shown in orange color). 
Next, we constrain the \pname{} cache to 8GB and measure the total execution time.
Storage I/O time (shown in grey color) can be computed by subtracting \texttt{Compute} and cache API time from this total time.

\reb{From Figure~\ref{fig:graphoverall}, 
we make the following key observations.  
First, the cache overhead is about 2-15\% for a single SSD and goes to 4-45\% for four SSDs. 
With a single SSD, the application performance is bounded by the storage throughput, and additional SSDs help to alleviate this problem by improving bandwidth, thus significantly minimizing the storage access overhead. 
The rest of the cache overhead is from cache metadata contention, long latency atomic operation, and warp scheduling in the face of polling threads. 
Even with this overhead, \pname{} outperforms the \textit{most optimistic baseline system available.}}
\reb{Second, with four SSDs, \pname{} is still bounded by the storage I/O throughput (5-6.2M IOPs which is >80\% of peak storage throughput), but there is an upper limit beyond which scaling SSDs will not help. This limit is bounded by the I/O request generation rate and how efficiently the applications utilize the cache. 
Further performance improvement requires application modification in work assignment/scheduling to 
trigger the \pname{} cache misses earlier during execution.}

\reb{\textbf{Sources of performance improvement:} Figure~\ref{fig:sourceperf} shows the sources of performance improvement with \pname{}. Using a naive cache (without warp coalescing or cache-line reference reuse) in \pname{} provides on average 11.9$\times$ (BFS) and 12.65$\times$ (CC) speedup over a no-cache implementation. This is because the \pname{} cache minimizes I/O amplification. When the application  exploits \pname{}'s warp coalescing and reference reuse efficiently, we observe an additional 6.07$\times$ (BFS) and 11.24$\times$ (CC) speedup on average.}

\begin{figure}[t]
\centering
{
  \includegraphics[width=\columnwidth]{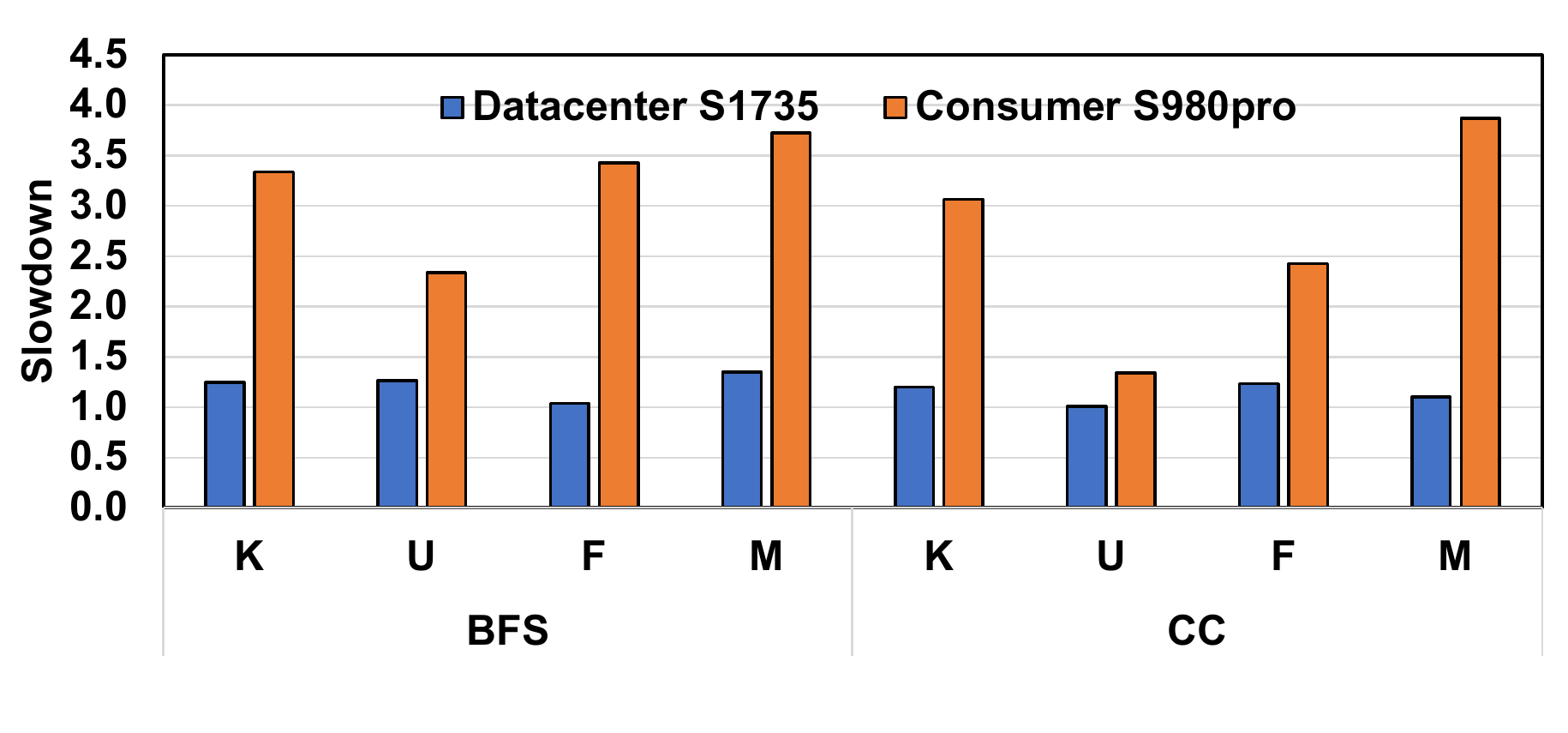}
}
 \vspace{-5ex}
\caption{The slowdown observed by \pname{} with four Samsung DC PM1735 (S1735) and Samsung 980pro SSDs when compared with four Intel Optane SSDs. } 
\label{fig:ssdcompare}
\vspace{-3ex}
\end{figure}

\textbf{Impact of SSD type:}
We now evaluate \pname{} with different types of SSDs: datacenter grade Samsung DC 1735 and consumer grade Samsung 980pro. 
\modi{Figure~\ref{fig:ssdcompare} shows the slowdown observed by \pname{} with each type using four SSDs when compared to Optane drives.}
Samsung DC 1735 and the Intel Optane SSD have similar performance for almost all workloads, since both achieve similar peak 4KB read IOPs.  
With the Samsung 980pro SSD, \pname{} prototype is on average 3.21$\times$ and 2.68$\times$ slower for BFS and CC workload. 
These results are very encouraging as the consumer-grade SSDs provide by far the \modi{(lowest cost)} among all the SSD technologies. 

\reb{
\textbf{Sensitivity Analysis:}
Next, we evaluate the effect of varying the cache size and the number of I/O queue pairs on the application performance with \pname{}.
We limit our discussions to the \texttt{K} dataset 
as similar trends are observed for other datasets.
}

\begin{figure}[t]
\centering
{
  \includegraphics[width=\columnwidth]{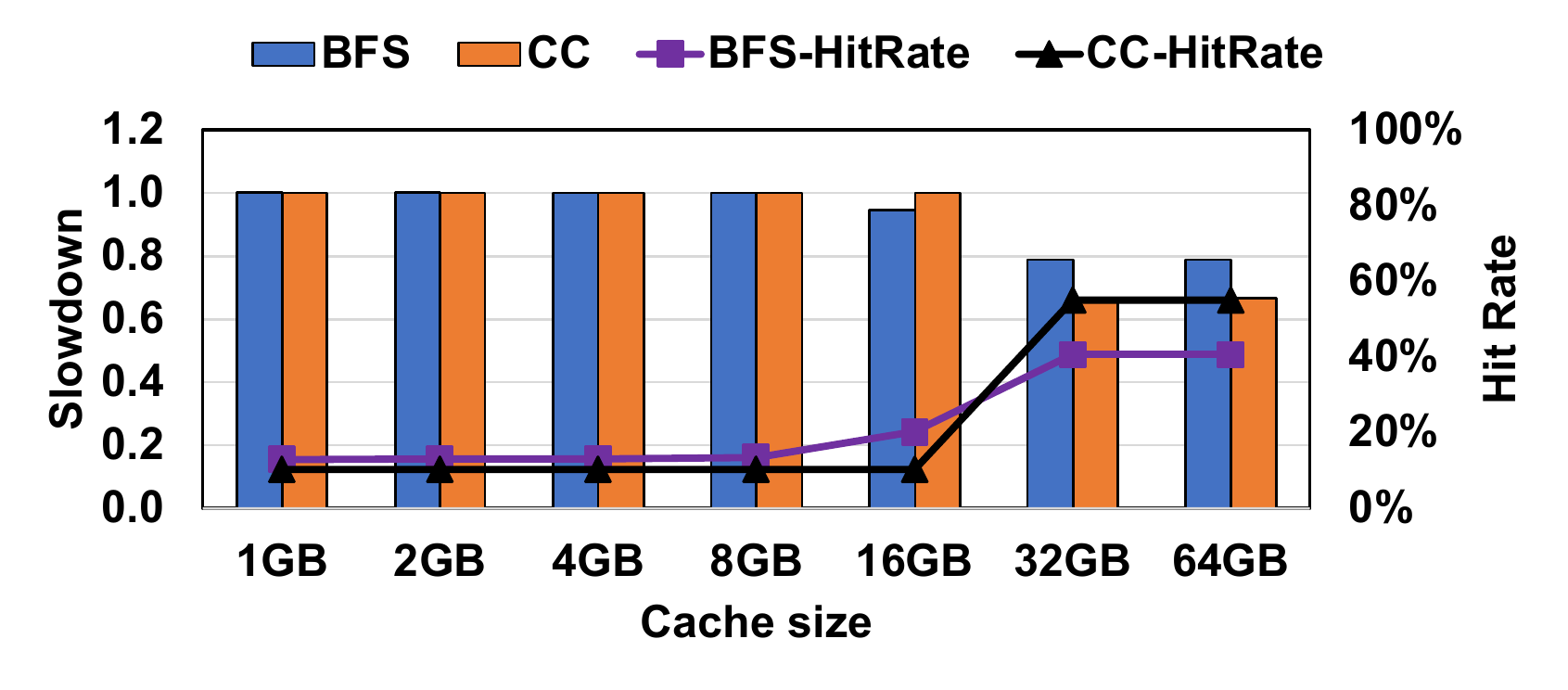}
}
 \vspace{-4ex}
\caption{\reb{Impact of \pname{} cache capacity for \texttt{K} dataset relative to an 8GB cache. \pname{} maintains the same performance with a 1GB cache as with a 8GB cache.}} 
\label{fig:cachecap}
\end{figure}

\reb{
\textit{Cache Size:}
As shown in Figure~\ref{fig:cachecap}, even with a 1GB cache, \pname{} does not experience any performance degradation as the cache can still capture the same level of spatial and temporal locality as 8GB.
Increasing the cache size to 32GB and 64GB enables the \pname{} cache to keep the entire working-set and only incur the cold cache misses.
}

\begin{figure}[t]
\centering
{
  \includegraphics[width=\columnwidth]{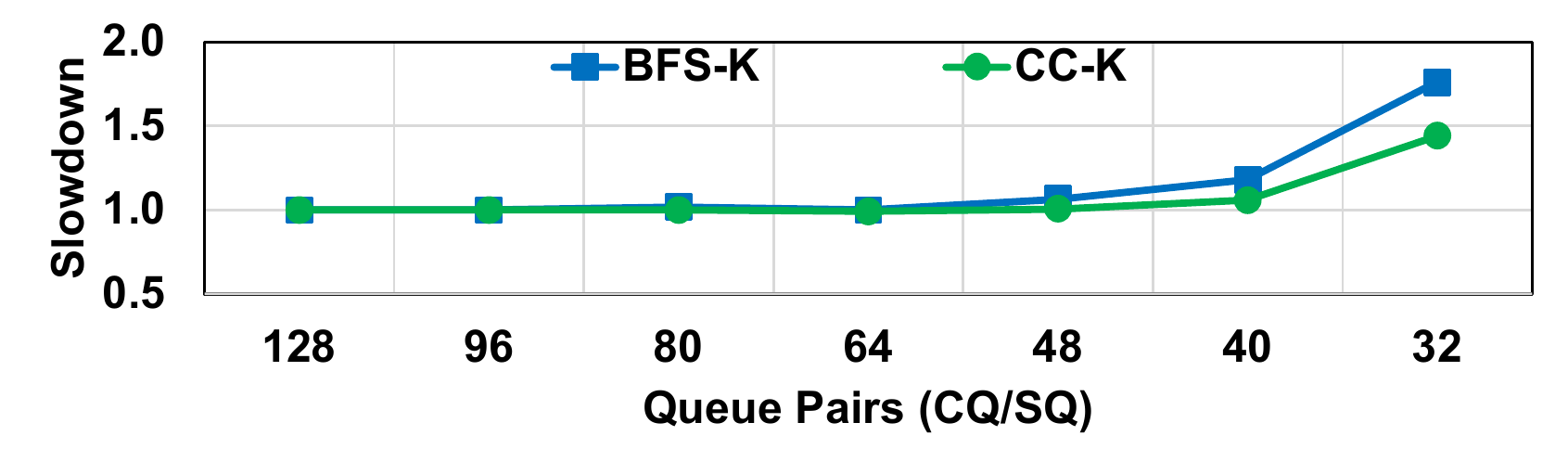}
}
 \vspace{-5ex}
\caption{\reb{Impact of the number of NVMe SQ/CQ queue pairs on performance for \texttt{K} dataset relative to 128 queue pairs. \pname{} maintains the same performance as the number of queue pairs decreases and only starts degrading with 40 queue pairs.}}
\label{fig:qpvary}
\end{figure}

\reb{
\textit{Number of Queue Pairs:}
As shown in Figure~\ref{fig:qpvary}, the application performance holds up well as the number of queue pairs decreases and degrades only at 40 (or lower) queue pairs, 
when the queue contention and the NVMe protocol serialization required per queue start to impact performance.
}
%

\subsection{I/O Amplification Benefit for Data Analytics} 
\label{sec:eval:ioamp}
Next, we evaluate the performance benefit of {the}  \pname{} prototype for enterprise data analytics workloads to illustrate \pname{}'s {benefits in reduced} I/O amplification and reduced software overhead 
while working on large structured datasets.
These emerging data analytics are widely used to interpret, discover or recommend meaningful patterns in data that is collected over time.
{Although} data-analytics applications are currently a small subset of all GPU applications, it is worth noting that the market size for this workload was \$205 billion in 2020 and \$230 billion in 2021~\cite{dataanalyticstam}.  

\textbf{Setup:} 
For this evaluation we use the NYC taxi ride dataset \cite{nycdataset} 
and six 
queries, to compare the performance of \pname{} against the state-of-the-art GPU accelerated data analytics framework, RAPIDSv21.12~\cite{rapids}. 
\modi{
The queries used for evaluation altogether answer the final query: 
``Q5: What is the average dollar/mile the driver makes for trips that are at least 30 miles?”
We start with a scan of the distance column (Q0), and add the total cost (Q1), surcharges (Q2), hail fee (Q3), tolls (Q4), and taxes (Q5) for only the trips that are at least 30 miles 
to get the penultimate query.
}
\modi{For RAPIDS, we pin the entire dataset file in the Linux CPU page cache, enabling RAPIDS to read the file contents directly from the CPU DRAM without issuing an I/O request to the storage. This captures the best performance modern systems can achieve.
For evaluating \pname{}, we replicate data across SSDs, using up to four Intel Optane P5800X SSDs with 4KB cache-lines and 8GB cache capacity. }

\begin{figure}[t]
\centering
{
  \includegraphics[width=\columnwidth]{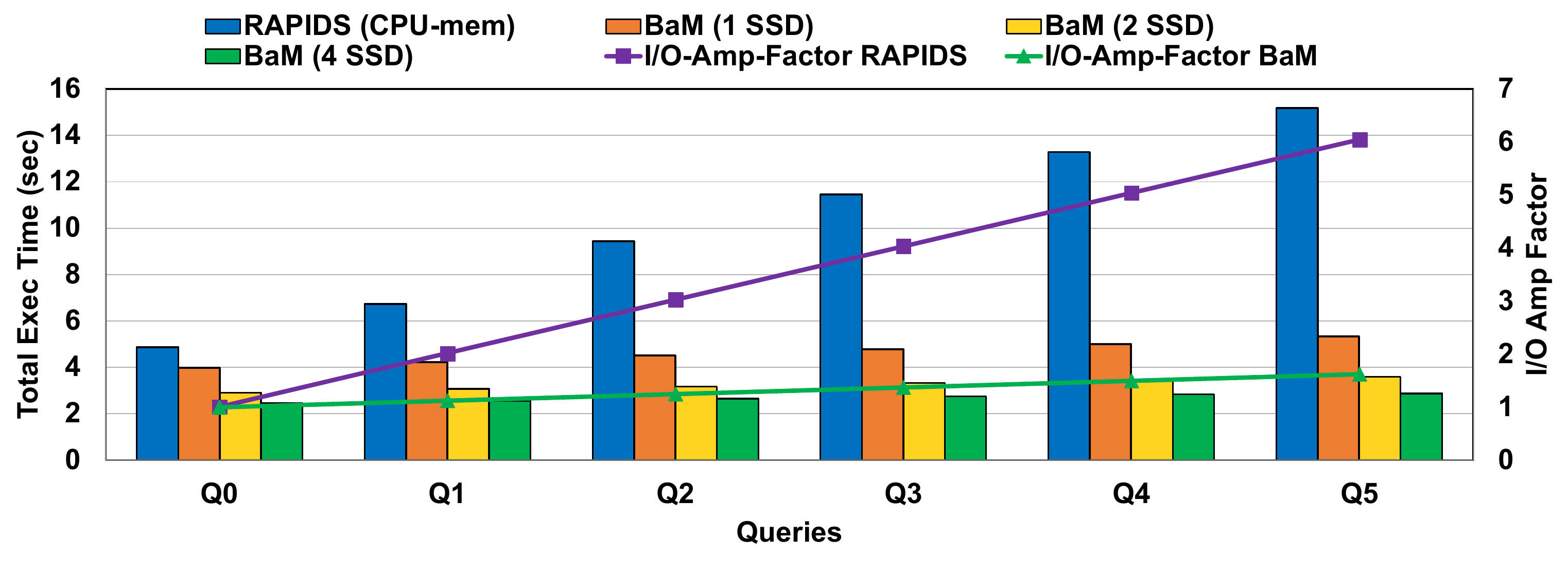}
}
\vspace{-4ex}
\caption{\reb{Performance of \pname{} and RAPIDS for data analytics queries on NYC Taxi dataset. \pname{} is up to 5.3$\times$ faster than the RAPIDS framework due to the reduced I/O amplification and reduced overhead in managing GPU memory.}  
}
\label{fig:cudfperf}
 \vspace{-3ex}
\end{figure}

\textbf{Results:} 
\modi{Even with the single SSD, \pname{} outperforms RAPIDS performance for all queries as shown in Figure~\ref{fig:cudfperf}.
For \texttt{Q0}, \pname{} observes a speed up of 1.22$\times$ over baseline even without any I/O amplification benefit. 
This is because, despite having the entire dataset preloaded into the CPU page cache, baseline RAPIDS experiences software overheads on the CPU to find and move data and manage the GPU memory.
}

\modi{With each additional data-dependent metrics (Q1 through Q5), the \pname{} performance advantage over RAPIDS increases as shown in Figure~\ref{fig:cudfperf}.}
The additional performance gain is attributed to \pname{}'s reduced I/O amplification due to on-demand data fetching, while 
RAPIDS must transfer entire columns to the GPU memory. 
With additional data-dependent metrics, as shown in Figure~\ref{fig:cudfperf}, the baseline suffers from 
{increased I/O amplification.} 
In contrast, \pname{}'s ability to make on-demand access to data as well as overlap compute, cache management, and many I/O requests helps it to handle multiple data-dependent columns nearly as efficiently as 
a single data-dependent column.

\modi{\pname{}'s end-to-end application time scales by up to 1.46$\times$ with two SSDs and 1.62$\times$ with four SSDs when compared to a single SSD \pname{} configuration. 
The sub-linear scaling is due to \pname{}'s setup overhead for pinning and mapping I/O queues and I/O buffers for DMA in GPU memory becomes more significant as more SSD's are used.  
Nevertheless, with four SSDs, \pname{} achieves up to 5.3$\times$ speed-up over the baseline. }


\reb{\subsection{VectorAdd Workload}}
\reb{In this section, we evaluate \pname{} on vectorAdd, 
a write-intensive workload. 
The vectorAdd workload takes two input arrays with four billion elements each, where each element is of eight-byte size to generate one output array of four billion elements. 
We assume the input vectors are in storage, and output requires to be written to the storage. 
For the baseline, 
we use a proactive tiling approach and split the four billion elements into five tiles.  
This allows the baseline to fully overlap the output write operation of the current tile with the loading of input vectors for the next tile. 
For \pname{}, 
the GPU kernel works on an entire dataset (no tiling), where each warp is assigned to a cache-line of the output vector. 
}

\reb{\pname{} is 1.51$\times$ slower 
than the baseline implementation. 
\pname{} currently does not support overlapping between read miss handling and write-back activities, thus exposing the entire write latency to the application. 
We can address this by enabling asynchronous write-back in the \pname{} system, 
which we leave as future work. }



\begin{figure}[t]
\centering
{
  \includegraphics[width=\columnwidth]{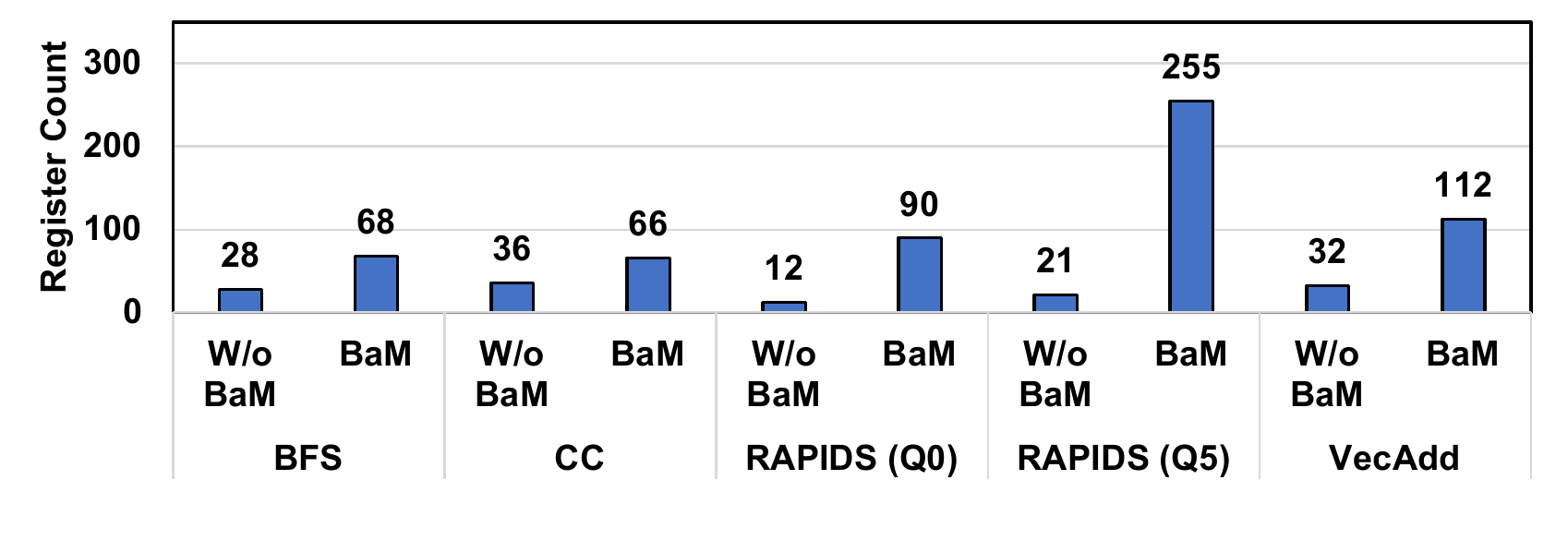}
}
 \vspace{-5ex}
\caption{\reb{Per-thread register usage in applications. Even though \pname{} requires more registers per thread, all studied applications are storage I/O bound and register spilling or the reduced occupancy does not bottleneck  performance.}}
\label{fig:reguse}
\vspace{-2ex}
\end{figure}
\reb{
\subsection{SM Resource Utilization}
Figure~\ref{fig:reguse} shows the register usage per thread with and without \pname{} in all studied applications. Register spilling is observed in the  RAPIDS workload. 
Neither \pname{} 
nor the applications 
use any shared memory. 
We do not enforce any occupancy constraints to limit register usage. 
}

\section{RELATED WORK}
\label{sec:related}

\subsection{Optimized CPU-Centric Model}
Most {existing} GPU programming models and applications 
were designed with the assumption 
that the working dataset fits in the GPU memory. 
If it doesn't, application specific techniques are employed to process large data on GPUs~\cite{zeroinfinity, dynamiccpugpu, gmac, gstream, gds, spin, morpheus, hippogriffdb, gullfoss, gpunet, gaia}. 

SPIN~\cite{spin} and NVMMU~\cite{nvmmu} propose to enable peer-to-peer (P2P) direct memory access using GPUDirect RDMA from SSD to GPU and exclude the CPU from the data path. 
SPIN integrates the P2P into the standard OS file stack and enables page cache and read-ahead schemes for sequential reads. 
GAIA~\cite{gaia} further extends SPIN's page cache from CPU to GPU memory. 
Gullfoss~\cite{gullfoss} provides a high-level interface that helps in initializing and using GPUDirect APIs efficiently. 
Hippogriffdb~\cite{hippogriffdb} provides P2P data transfer capabilities to the OLAP database system. 
GPUDirect Storage~\cite{gds} is a product that removes the CPU memory from the data path between SSDs and GPUs using GPUDirect RDMA technology. AMD has had similar efforts in RADEON-SSG products~\cite{ssg}.
These works still require the CPU to orchestrate data movement.
\pname{} allows any GPU thread to initiate data accesses to the SSDs. 

\subsection{Prior Accelerator-Centric Systems}
ActivePointers~\cite{activepointers}, GPUfs~\cite{gpufs},  GPUNet~\cite{gpunet}, and Syscalls for GPU~\cite{amdsyscall} have previously attempted to enable accelerator-centric model for data orchestration.
GPUfs~\cite{gpufs} and Syscalls for GPU~\cite{amdsyscall} first allowed GPUs to request file data from the host CPU.
ActivePointers~\cite{activepointers} added a memory-map like abstraction on top of GPUfs to allow GPU threads memory-like access to file data.
Dragon~\cite{dragon} incorporated storage access to the UVM~\cite{uvm} page faulting mechanism. 
However, as shown in $\S$~\ref{sec:eval} and ~\cite{bamarxiv}, these approaches use less-parallel CPUs to handle data demands from the massively parallel GPU and result in 
poor overall performance.

\reb{We also acknowledge the prior work in moving network control to the GPU~\cite{gpurdma}, however storage presents a new set of challenges.
With \pname{}, we enable GPUs to efficiently and directly access storage and show the performance and cost benefits for real applications.}


\subsection{Hardware Extensions}
Prior work has proposed to replace or integrate GPU's global memory with non-volatile memories~\cite{flashgpu, zng, nvmopps,ohmgpu,gpunvm}.
DCS~\cite{dcs} proposed enabling direct access between storage, network, and accelerators with an FPGA providing the required translation services for coarse-grain data transfers.
Enabling persistence within the GPU has recently been proposed~\cite{gpupersist}.
We acknowledge these efforts and further validate the need to enable large memory capacity for emerging workloads.  
More importantly, the \pname{} prototype aims to use 
{emerging disaggregated storage} hardware {components} to provide significant performance advantages to end-to-end applications with very large real-world datasets. 

\section{CONCLUSION}
\label{sec:conclusion}

In this work, we make a case for enabling GPUs to orchestrate high-throughput, fine-grain{ed}  accesses to storage\reb{, without the CPU {software overhead},} in a new system architecture called~\pname{}. 
\pname{} mitigates the I/O amplification problem 
by {allowing the GPU application compute code to read or write at finer granularities on-demand.} 
\modi{As \pname{} supports the storage access control plane functionalities including caching, translation, and protocol queues on the GPU, it avoids the costly CPU-GPU synchronization, OS kernel crossing, and software bottlenecks that have limited the achievable storage access throughput.}
Using off-the-shelf hardware components, we 
{built} a prototype of \pname{} and show on multiple applications and datasets that ~\pname{} is a viable/superior alternative to DRAM-only and other state-of-the-art solutions. 

\begin{acks}
We would like to acknowledge all of the help from members of {the} IMPACT research group, {the} IBM-{Illinois} Center for Cognitive Computing Systems Research (C3SR) and NVIDIA Research without which we could not have achieved any of the above reported results. 
Special thanks to Kun Wu from {the} IMPACT research group for his suggestion on using advanced warp primitives for implementing fast warp coalescing.
We would also like to acknowledge the valuable insight and help we received from stake holders including NVIDIA, Intel, Samsung, Phison, H3 Platform,
Broadcom, and University of Illinois. Especially discussions
with Yaniv Romem, Brian Pan, Jean Chou, Jeff Chang, Yt
Huang, Annie Foong, Andrzej Jakowski, Allison Goodman,
Venkatram Radhakrishnan, Max Simmons, Michael Reynolds,
John Rinehimer, In Dong Kim, Young Paik, Jaesung Jung,
Yang Seok Ki, Jeff Dodson, Raymond Chan, Hubertus Franke,
Paul Crumley, Sanjay Patel, Deming Chen, Josep Torrellas,
David A. Padua and several others have been fundamental
for this work to come to fruition. This work uses GPUs donated by NVIDIA and is partly supported by {the} IBM-ILLINOIS
C3SR and by the IBM-ILLINOIS Discovery Accelerator
Institute (IIDA).
\end{acks}


%
%
%
%
%


\appendix
\section{ARTIFACT APPENDIX}

\subsection{Abstract}
BaM 
{is} a novel system architecture that 
{defines} mechanisms for GPU threads to efficiently {initiate and orchestrate storage accesses.} 
As such, with the artifact{, i.e., a prototype implementation of BaM, the authors 
demonstrated the following:} 
(1) BaM is functional with a single SSD and provides the contributions described in the paper,
(2) BaM is functional with multiple SSDs - up to 2 SSD{s} can be tested 
{with} the provided system, 
(3) BaM is functional with different types of SSDs - consumer-grade Samsung 980 Pro and datacenter-grade Intel Optane SSDs available in the provided system, 
(4) BaM works with different applications - microbenchmarks and graph applications (with real datasets) {that} are provided for artifact evaluation. 
{Based on these demonstrations, 
this paper} has been awarded the Artifact Available and Artifact Functional badges.

\subsection{Artifact Checklist (Meta-information)}


{\small
\begin{itemize}
  \item {\bf Compilation:} For the artifact evaluation reviewers, the authors provided access to a machine where the required compilation tool chains was already set up. For general details, refer to software dependencies ($\S$\ref{sec:soft_dep}).
  \item {\bf Dataset:} As the dataset size is over 500GB, the datasets were pre-loaded on the SSDs in the provided system for the artifact evaluation reviewers. For general details refer to ($\S$\ref{sec:datasets}).
  \item {\bf Run-time environment:} Linux Kernel 5.8.x. CUDA, etc. Refer to software dependencies ($\S$\ref{sec:soft_dep}) for details. Root access is required.
  \item {\bf Hardware:} For the artifact evaluation reviewers, the authors provided access to a machine where the required hardware was already set up. For general details refer to hardware dependencies ($\S$\ref{sec:hard_dep}).
  \item {\bf Run-time state:} Yes
  \item {\bf Execution:} Only one artifact evaluation reviewer can execute the evaluation applications at a time on the provided machine. 2-3 hours aggregate run time.
  \item {\bf Metrics:} Execution time, bandwidth and IOPS.
  \item {\bf Output:} Console. See provided log files for expected output.
  \item {\bf Experiments:} Follow the instructions at \url{https://github.com/ZaidQureshi/bam/tree/master/asplosaoe#readme}
  \item {\bf How much disk space required (approximately)?:} >500GB. 
  \item {\bf How much time is needed to prepare workflow (approximately)?:} 30 minutes
  \item {\bf How much time is needed to complete experiments (approximately)?:} 3 hours
  \item {\bf Publicly available?:} Yes
  \item {\bf Code licenses (if publicly available)?:} BSD-2-Clause
  \item {\bf Archived at:} https://doi.org/10.5281/zenodo.7217356 \cite{zenodo}
\end{itemize}
}

\subsection{Description}

\subsubsection{Access}

\pname{}'s codebase is publicly available here: \url{https://github.com/ZaidQureshi/bam.git}

\subsubsection{Hardware dependencies}
\label{sec:hard_dep}
Refer to \url{https://github.com/ZaidQureshi/bam#hardwaresystem-requirements}

\subsubsection{Software dependencies}
\label{sec:soft_dep}
Refer to \url{https://github.com/ZaidQureshi/bam#system-configurations}

\subsubsection{Datasets}
\label{sec:datasets}
Refer to \url{https://github.com/ZaidQureshi/bam#example-applications}

\subsection{Installation}
The complete installation instructions can be found at the following link \url{https://github.com/ZaidQureshi/bam/blob/master/asplosaoe/README.md#compiling}. We briefly describe them in this section.

\subsubsection{Building the Project}
From the project root directory, follow the steps shown in Listing~\ref{lst:buildbam}. 
The CMake configuration is supposed to auto-detect the location of CUDA, Nvidia driver and project library. 
After this, we should also compile the custom \texttt{libnvm} kernel module for NVMe devices as shown in the last two lines in Listing~\ref{lst:buildbam}. 

\begin{lstlisting}[basicstyle=\scriptsize,linewidth=8.3cm,float=t!,language=bash,caption={\small Building BaM library, applications and kernel module.}, label=lst:buildbam]
# Building the BaM library and applications
$ git submodule update --init --recursive
$ mkdir -p build; cd build
$ cmake ..
$ make libnvm -j     # builds library
$ make benchmarks -j # builds benchmark program

# Building the BaM kernel module
$ cd module
$ make -j
\end{lstlisting}

\subsubsection{Loading/Unloading the Kernel Module}
In order to use the custom kernel module for the NVMe device, we need to unbind the NVMe device from the Linux NVMe driver. To do this, first we need to find the PCIe ID of the NVMe device. 
If the required NVMe device want is mapped to the \texttt{/dev/nvme0} block device, Listing~\ref{lst:loaddriver} shows how to find its PCIe ID.

\begin{lstlisting}[basicstyle=\scriptsize,float=t!,linewidth=8.3cm,language=bash,caption={\small Identifying SSD to use and binding it to the BaM driver.}, label=lst:loaddriver]
$ dmesg | grep nvme0
[126.497670] nvme nvme0: pci function 0000:65:00.0
[126.715023] nvme nvme0: 40/0/0 default/read/poll queues
[126.720783] nvme0n1: p1
[190.369341] EXT4-fs (nvme0n1p1): ...

# Unbind the SSD from kernel NVMe driver
$ echo -n "0000:65:00.0" > 
   /sys/bus/pci/devices/0000\:65\:00.0/driver/unbind

# Load the BaM driver.
$ cd build/module
$ sudo make load
\end{lstlisting}

Next, we need to unbind the SSD from the Linux NVMe driver, as shown in Listing~\ref{lst:loaddriver}. Repeat the process for all NVMe devices that are to be remapped to the BaM driver. We next load the BaM kernel module from the build directory. 
This should create a \texttt{/dev/libnvm*} device file for each NVMe SSD that is not bound to the NVMe driver. 

\subsubsection{Basic Test}
\begin{lstlisting}[basicstyle=\scriptsize,float=t!,linewidth=8.3cm,breaklines=true,language=bash,caption={\small Basic test.}, label=lst:test]
$ sudo ./bin/nvm-block-bench --threads=262144 --blk_size=64 --reqs=1 --pages=262144 --queue_depth=1024  --page_size=512 --num_blks=2097152 --gpu=0 --n_ctrls=1 --num_queues=128 --random=true
\end{lstlisting}
The evaluator can run the command available at \url{https://github.com/ZaidQureshi/bam/blob/master/asplosaoe/README.md#running-the-io-stack-component} (also shown in Listing~\ref{lst:test}) as a basic test.
The expected output (sans performance numbers) can be found at \url{https://github.com/ZaidQureshi/bam/blob/master/asplosaoe/nvm_block_bench_1_sam.log}.


\subsection{Evaluation}
\subsubsection{Experimental workflow}
Each benchmark is an executable compiled in the process described earlier. Each benchmark has various command line parameters (explained in its help) and outputs relevant performance metrics like time and throughput.

\subsubsection{Evaluation and expected results}
Evaluation should validate the goals that are defined in the artifact evaluation abstract. For each of the goals, we have prepared a set of commands,
their descriptions and expected results at  \url{https://github.com/ZaidQureshi/bam/tree/master/asplosaoe#readme}.








\section{Background and Motivation}
\label{apdx:background}

This is extended $\S$\ref{sec:background} version covering additional details to motivate the design of \pname{}. 
State-of-the-art CPU-centric solutions  
differ in whether the application CPU software \textit{proactively} loads the data from storage into GPU memory before any GPU computation (See $\S$~\ref{apdx:cudf}) or system software \textit{reactively} accesses the storage to service page faults from the GPU (See $\$$~\ref{apdx:faults}).

\subsection{Proactive Tiling}
\label{apdx:cudf}
Proactive tiling is a CPU-centric solution that 
requires the programmer to explicitly decompose and partition the data into tiles that fit into the GPU memory.
The CPU application code orchestrates data movement between the storage and the GPU memory to \textit{proactively} preload tiles into GPU memory, launches compute kernels for each tile, and aggregates the results from processing the individual tiles.
Although, proactive tiling works well for some classical GPU applications with predefined, regular, and dense access patterns, its proactive accesses to storage are problematic for emerging applications with dynamic, { data-dependent, irregular access patterns}, such as data analytics.
The execution time overhead of the synchronization and CPU orchestration compels the developers to resort to coarse-grain tiles, which exacerbates I/O amplification. 

\begin{figure}[t]
\centering
{
  \includegraphics[width=\columnwidth]{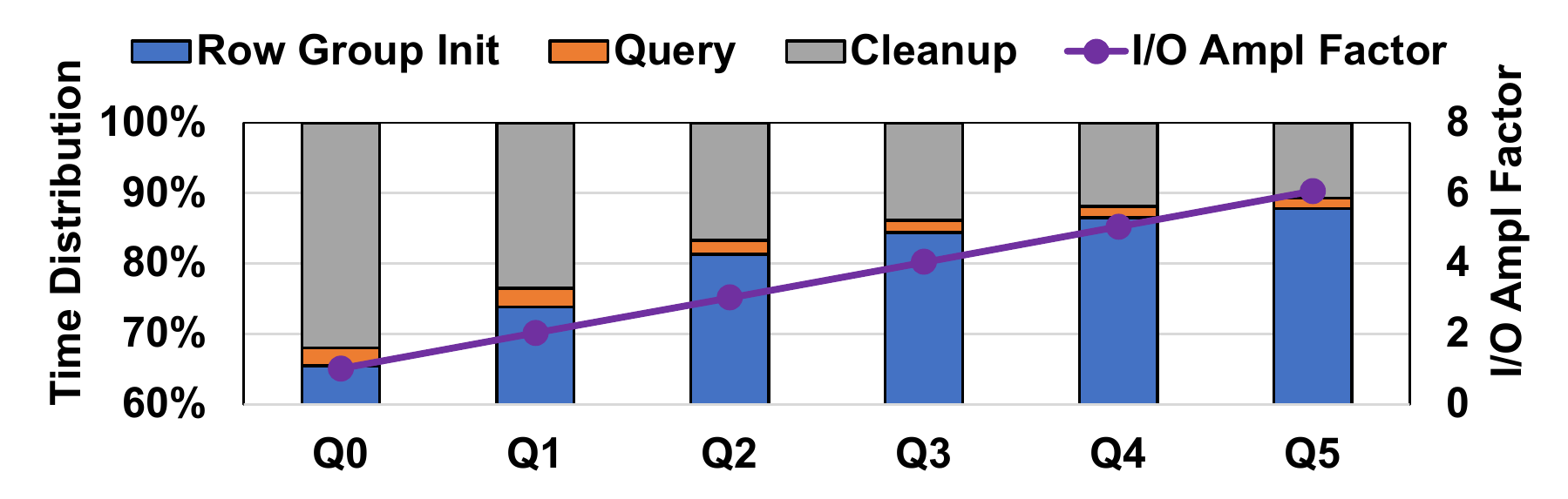}
}
\vspace{-4ex}
\caption{ Execution time breakdown and I/O amplification in GPU-accelerated data analytics application using the state-of-the-art RAPIDS~\cite{rapids} system.
} 
\label{fig:cudf}
\end{figure}

Take executing analytics queries on the NYC taxi ride dataset (See $\S$\ref{sec:eval:ioamp} for details) as an example. {The NYC Taxi dataset consists of 1.7 billion taxi trip records from 2009 to 2021. It is organized as a table where each row is a trip record and each column stores a metric such as pickup location, distance, total fare amount, taxes, surcharge, etc. }
Suppose we ask the question, ``Q1: What is the average cost/mile for trips that are at least 30 miles?”.  
As the data-set has almost 2 billion rows and exceeds the GPU memory capacity, the programmer must process the data in smaller row groups that fit in GPU memory by (1) finding and loading each row group {(i.e., tile)} into the GPU memory, (2) performing the query on the GPU over the row group, and (3) aggregating results across row groups.
During query computation over the row group, the trip distance column of the row group is scanned, and, for trips that meet the criteria of being at least 30 miles, the total cost value of the corresponding row must be aggregated.
Figure~\ref{fig:cudf} shows the profiling result for executing this query on the state-of-the-art GPU accelerated data analytics framework, RAPIDSv21.12~\cite{rapids}.
The CPU code to initialize the row group, which involves finding, {allocating memory for} and loading each row group {of the columnar metric arrays} into GPU memory, and the CPU code to clean up the row group {accounts for more than 73\% and 23\%, respectively,} of the end-to-end application time, {reflecting the high driver and synchronization overhead}


Furthermore, since accesses to the total cost column are dependent on values in the trip distance column, the CPU cannot determine which total cost rows are required. 
Thus, to leverage the bandwidth of the storage, RAPIDS fetches all rows of both columns into the GPU's memory for each row group.
As only 511k trips are at least 30 miles, only 0.03\% of the second column is used. 
Thus, the tiling approach incurs 2.02$\times$ I/O traffic amplification for this query. 


The query can be further extended to answer a more interesting question: ``Q5: What is the average \$/mile the driver makes for trips that are at least 30 miles?".
To answer this query we need 4 more data-dependent metrics from the data-set and for each metric added, we create a new intermediate query.
We have to add the surcharges (Q2), hail fee (Q3), tolls (Q4), and taxes (Q5) metrics to get the penultimate query. 
But doing so \textit{linearly scales the I/O amplification suffered by the CPU-centric model {to over 6$\times$}} as the number of data-dependent metrics increases, as shown in Figure~\ref{fig:cudf}.


\begin{figure}[t]
\centering
{
  \includegraphics[width=\columnwidth]{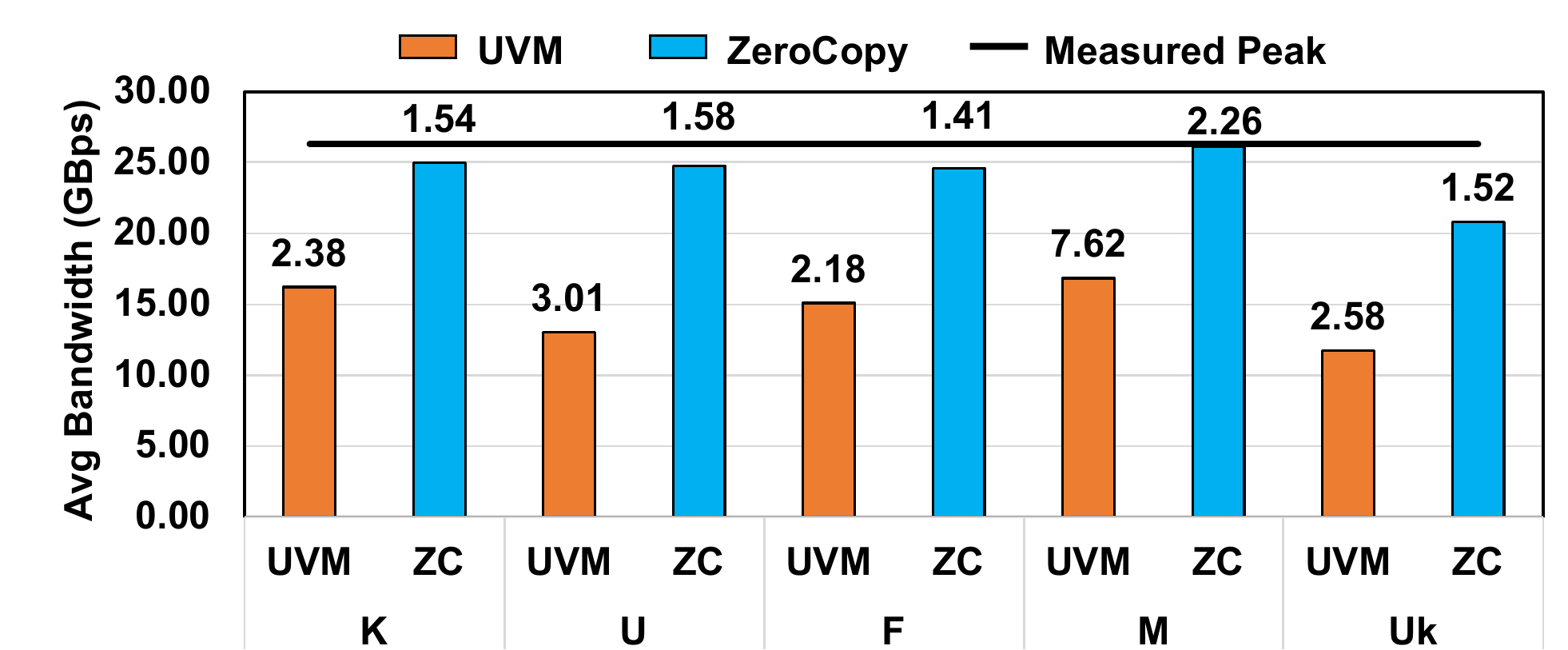}
}
\vspace{-2ex}
\caption{Average bandwidth in GBps (bar graph) and execution time in seconds (numbers) when running BFS graph traversal using UVM and Zerocopy mechanism. UVM page faulting scheme adds significant overhead.
} 
\label{fig:softuvm}
\end{figure}
 
\subsection{Reactive Page Faults}
\label{apdx:faults}
{Some applications, such as graph traversal, do not have ways to cleanly partition their datasets and thus prefer keeping whole data structures in the GPU's address space.}
For example, assume that a graph is represented in the popular compressed sparse row (CSR) format, where the neighbor-lists of all nodes are concatenated into one large edge-list array.
An array of offsets accompanies the edge-list, where the value at index \texttt{i} specifies the starting offset for the neighbor-list of node \texttt{i} in the large edge-list.
Since any node, and its corresponding neighbor-list, can be visited while traversing a graph, {traversal algorithms} prefer to keep the entire edge-list in the GPU's address space~\cite{emogi}.

Starting with the NVIDIA Pascal architecture, GPU drivers and programming model allow the GPU threads to 
access large virtual memory objects that may partly reside in the host memory using Unified Virtual Memory (UVM) abstraction~\cite{uvm}.
Prior work shows that the UVM driver can be extended to interface with the file system layer to access memory-mapped files
~\cite{dragon}.
This enables a GPU to generate a page fault for data not in GPU memory which the UVM driver \textit{reactively} services by making I/O request(s) for the requested pages on storage.

However, this 
approach introduces significant software overhead in 
page fault handling mechanisms when the accessed data is missing from GPU memory. 
Figure~\ref{fig:softuvm} shows the measured host-memory-to-GPU-memory data transfer bandwidth 
for an A100 GPU in a PCIe Gen4 system executing BFS graph traversal on six different datasets (See Table~\ref{tab:graphdataset}) where the edgelists are in the UVM address space and initially in the host memory. {Note that there is no storage access in this experiment and the measured data bandwidth is an upper-bound of what a page fault-based approach might potentially achieve.}

From Figure~\ref{fig:softuvm}, the average PCIe data transfer bandwidth achieved by the UVM page faulting mechanism is $\sim$14.52GBps which is only 55.2\% of the measured peak PCIe Gen4 bandwidth.
Profiling data 
shows that the UVM fault handler on the CPU is 100\% utilized and the maximum UVM page fault handling rate saturates at $\sim$500K IOPs. 
Such a rate is only half of the peak throughput of a Samsung 980pro SSDs 
as noted in Table~\ref{tab:ssdstat}.
With these limitations, 
the UVM page fault handling mechanism simply cannot 
fully utilize the bandwidth of even one consumer-grade SSD for page sizes of 8K or smaller.

\balance
\bibliographystyle{ACM-Reference-Format}
\bibliography{refs}

\end{document}